\documentclass[a4paper,11pt]{article}
\usepackage{graphics}
\usepackage{amsmath}
\usepackage{amssymb}
\usepackage{tikz-cd}
\usepackage{slashed}
\begin{document}
\title{\textbf{Neutrino oscillations and  Lorentz Invariance Violation in a Finslerian Geometrical model}}
\author{V. Antonelli, L. Miramonti, M.D.C. Torri\\
Dipartimento di Fisica,\\
Universit\`a degli Studi di Milano e INFN\\
Via Celoria 16, 20133 - Milano%
}     
%
%
%

\maketitle

\begin{abstract}
{Neutrino oscillations are one of the first evidences of physics beyond the Standard Model (SM). Since Lorentz Invariance is a fundamental symmetry of the SM, recently also neutrino physics has been explored to verify the eventual modification of this symmetry and its potential magnitude. In this work we study the consequences of the introduction of Lorentz Invariance Violation (LIV) in the high energy neutrinos propagation and evaluate the impact of this eventual violation on the oscillation predictions. An effective theory explaining these physical effects is introduced via Modified Dispersion Relations. This approach, originally introduced by Coleman and Glashow, corresponds in our model to a modification of the special relativity geometry. Moreover, the generalization of this perspective leads to the introduction of a maximum attainable velocity which is specific of the particle. This can be formalized in Finsler geometry, a more general theory of space-time. In the present paper the impact of this kind of LIV  on neutrino phenomenology  is studied, in particular by analyzing the corrections introduced in neutrino oscillation probabilities for different values of neutrino energies and baselines of experimental interest. The possibility of further improving the present constraints on CPT-even LIV coefficients by means of our analysis is also discussed.}
\end{abstract}
\section{Introduction}
\label{sec:1}
Till now it is not possible to reach sufficiently high energies to probe the structure of space-time at Planck scale, which is considered the separation point of standard gravitational theories from quantized ones. Nevertheless, space-time quantum effects can possibly manifest as little deviations from the \emph{standard physics} predictions. Hence, even if there are no definitive evidences to sustain departures from Lorentz Invariance, it is possible that this symmetry emerges at ``low" energies as an effective symmetry, but is violated in a more energetic scenario, when the quantum effects start to be recognizable. Experimental observations, conducted on the propagation of high energy cosmic messengers, hint at the possibility that their propagation could be influenced by some deviations from the standard physical theories~\cite{Tests-LIV}. 

Coleman and Glashow~\cite{Glashow1} were the first to introduce the
hypothesis of Lorentz Invariance Violation (LIV), as an attempt to justify such experimental observations and to explore some consequences even in the neutrino sector. Other works~\cite{Glashow2,Liberati1}
dealt with the effects of LIV on neutrino physics, but they focused on posing constraints on the maximum magnitude of perturbations for ultra-luminal neutrinos or investigated the possibility that the masses of neutrinos are generated in a modified relativity scenario. In the model we are going to discuss we consider, instead, Lorentz invariance violating effects as tiny deviations, that affect the oscillation sector without modifying the general pattern.

The existence of neutrino oscillations itself violates the original Standard Model predictions and seems, therefore, to require the introduction of new physical theories, beyond the ``minimal version'' of the Standard Model (in which neutrinos would be simply left handed massless Dirac fermions).
Also for this reason, it is very interesting to explore the phenomenology introduced by LIV on very energetic particles, even in neutrino oscillation sector. In this work we introduce the Lorentz symmetry violation from Modified Dispersion Relations (MDRs), assumed as consequence of an underlying more general relativity theory, that modifies the kinematics. From this starting point, we show the need to resort to Finsler geometry~\cite{Finsler}, to construct an effective geometrical theory, which can account for LIV perturbations.

Finally, we explore the phenomenological  consequences introduced in the neutrino oscillation physics
by the presence of LIV violating corrections and by the consequent modifications of dispersion relations. We focus in particular on the analysis of the way in which the oscillation probabilities, which rule the neutrino flavor transitions, get modified for different values of neutrino energies and baselines, with particular attention to the values relevant for long-baseline accelerator and atmospheric neutrinos and for high energy cosmic neutrinos.
We also discuss the way in which our analysis can be compared with similar studies developed in literature
(even if in different kind of models in most cases) and
the possibility of imposing more severe constraints on the LIV coefficients with a similar analysis applied to future neutrino experiments.
\section{Modified Dispersion Relations introduced LIV and Finsler Geometry}
\label{sec:2}

One simple way to introduce LIV consists in modifying the kinematics of the theory, that is the Dispersion Relations (DR). As shown in~\cite{Glashow1}, imposing a maximum speed,
 lower than speed of light, for a massive particle implies a modification of the DR, given by $E^2=(1-\epsilon)^2|\overrightarrow{p}|^2+(1-\epsilon)^4m^2$, which,
reabsorbing the negligible correction term proportional to the mass, can be written as:
\begin{equation}
\label{mdr1}
E^2-(1-\epsilon)^2|\overrightarrow{p}|^2=m^2 \, .
\end{equation}
In the previous equations $\epsilon\ll1$ indicates a constant, representing the maximum speed modification parameter.\\

Following the original idea proposed in~\cite{Maguejo}, this type of MDR can be generalized in a form, which includes energy dependent corrections:
\begin{equation}
\label{mdr2}
f_{1}^{2}E^2-f_{2}^{2} p^2=m^2\, ,
\end{equation}
where $f_{i}$ are functions of the quadrimomentum p, that can be written as $f_{i}=1-h_{i}$ and $h_{i}\ll1$ are the velocity modification parameters. From this relation, it is possible to derive an explicit equality for the energy:
\begin{equation}
\label{energy}
E=\sqrt{\frac{m^2}{f_{1}^{2}}+\frac{f_{2}^{2}}{f_{1}^{2}}\, p^2}\simeq p f_{3}\, \, , with \,  f_{3}=\frac{f_{2}}{f_{1}} \, .
\end{equation}

Using the Hamilton-Jacobi equation, one gets for the velocity:
\begin{equation}
\label{hamilton1}
\frac{\partial}{\partial p}E=f_{3}+ p \, f'_{3} \, .
\end{equation}
This means that every propagating lepton feels a local space-time foliation, parameterized by its momentum, that is by its energy. From this the necessity follows to resort to a more general geometry, that can account for this energy dependence, the Finsler geometry~\cite{Finsler}.\\
Following the work of~\cite{Torri}, we introduce the LIV, even in neutrino sector, via the Modified Dispersion Relations (MDRs) with the form:
\begin{equation}
\label{mdr3}
E^2-\left(1-f\left(\frac{|\overrightarrow{p}|}{E}\right)-g\left(\frac{\overrightarrow{p}}{E}\right)\right)|\overrightarrow{p}|^2=m^2
\end{equation}
where the perturbation functions:
\begin{equation}
\label{pert}
\begin{split}
&f\left(\frac{|\overrightarrow{p}|}{E}\right)=\sum_{k=1}^{\infty}\alpha_{k}\left(\frac{|\overrightarrow{p}|}{E}\right)^k\\ &g\left(\frac{\overrightarrow{p}}{E}\right)=\sum_{k=1}^{\infty}\beta_{k}\left(\frac{\overrightarrow{p}}{E}\right)^k
\end{split}
\end{equation}
are chosen homogeneous in order to guarantee the geometrical origin of the MDR, as it happens in special relativity, where the Dispersion Relations are written using the Minkowski metric as: $E^2-|\overrightarrow{p}|^2=m^2$ $\Rightarrow\;\eta^{\mu\nu}p_{\mu}p_{\nu}=m^2$.

It is important to underline that the perturbation $f$ preserves the isotropy of space; this is not true, instead, for the function $g$, that introduces a preferred direction. Therefore, the form of the MDR can be chosen in such a way to preserve, or not, the idea of a privileged frame of reference. In this work, for simplicity, we assume the space to be isotropic, posing $g=0$, but all the results are still valid even in the other case.

Moreover, it is important to be cautious in defining the adimensional
coefficients $\alpha$ and $\beta$ in (\ref{pert}), to guarantee that the energy, as function of the momentum, assumes positive finite values. In this way the ratio
$\frac{|\overrightarrow{p}|}{E}\rightarrow1+\delta$ admits a limit for $p\rightarrow\infty$ and, consequently, even the perturbation functions admit limit,
$f(1+\delta)=\epsilon$ and $g(1+\delta)=\epsilon$, if not posed equal to zero. In this way it is possible to reobtain the Coleman and Glashow Very Special Relativity (VRS) scenario, with the
perturbation function $f_3$ that, for $p\rightarrow\infty$, tends to
$${ \lim_{p\rightarrow\infty} \, f_3 = 1- f(1+\delta) = 1-\epsilon}\, .$$
Hence, from eq.(\ref{hamilton1}), one recovers
for the ``personal'' \emph{maximum attainable velocity} $c'$ of a massive particle, the constant value different from the light speed:
\begin{equation}
\label{maxattvel}
c'=\frac{\partial}{\partial p}E\bigg\vert_{max}=f_{3} = 1-\epsilon \, .
\end{equation}

The hypothesis made on the perturbation functions permit to write the MDRs (\ref{mdr3}) as:
\begin{equation}
\label{mdr4}
\widetilde{g}(p)^{\mu\nu}p_{\mu}p_{\nu}=F^2(p)=m^2
\end{equation}
and using the equation:
\begin{equation}
\label{metric0}
\widetilde{g}(p)^{\mu\nu}=\frac{1}{2}\frac{\partial}{\partial p^{\mu}}\frac{\partial}{\partial p^{\nu}}F^2(E,\,\overrightarrow{p})
\end{equation}
we obtain the explicit form of the metric, defined in the momentum space, after eliminating a non-diagonal part, that gives no contributions in computing the dispersion relation:
\begin{equation}
\label{metric1}
\widetilde{g}(p)^{\mu\nu}=\left(
                         \begin{array}{cccc}
                           1 & 0 & 0 & 0 \\
                           0 & -(1-f(p)) & 0 & 0 \\
                           0 & 0 & -(1-f(p)) & 0 \\
                           0 & 0 & 0 & -(1-f(p)) \\
                         \end{array}
                       \right)
\end{equation}

The homogeneity of the perturbation $f$ implies that the function $F$ defined in (\ref{mdr4}) is homogeneous of degree $1$, condition to be a Finsler norm; hence, even the derived metric is defined in a Finsler space. The properties of this geometry allows to define the \emph{Legendre} transformation of the metric, as a bijection, to obtain the corresponding tensor in coordinate space and results $g_{\mu\nu}(x)=\widetilde{g}_{\mu\nu}(p)$. Therefore, we 
obtain the generic metric depending both on coordinates and momentum:
\begin{equation}
\label{metric2}
g(x,\,p)_{\mu\nu}=\left(
                         \begin{array}{cccc}
                           1 & 0 & 0 & 0 \\
                           0 & -(1+f(p)) & 0 & 0 \\
                           0 & 0 & -(1+f(p)) & 0 \\
                           0 & 0 & 0 & -(1+f(p)) \\
                         \end{array}
                       \right)
\end{equation}

\section{More on the geometry of space-time}
\label{sec:3}
In order to have a deeper insight in the introduced geometrical structure, we have to deal with the foliation of the space-time depending on the momentum magnitude. To do this it is necessary to introduce the Cartan formalism and, therefore, to resort to the \emph{vierbein} or \emph{tetrad}, whose form is given by:
\begin{equation}
\label{vierbein}
\begin{split}
&\left[e\right]_{\mu}^{\,a}=\left(
                             \begin{array}{cc}
                               1 & \overrightarrow{0} \\
                               \overrightarrow{0}^{t} & \sqrt{1+f(p)}\,\mathbb{I} \\
                             \end{array}
                           \right)\\
&\left[e\right]^{\mu}_{\,a}=\left(
                             \begin{array}{cc}
                               1 & \overrightarrow{0} \\
                               \overrightarrow{0}^{t} & \sqrt{1-f(p)}\,\mathbb{I} \\
                             \end{array}
                           \right)
\end{split}
\end{equation}
where the dependence on the momentum is evident. Using the \emph{tetrad}, it is possible to evaluate the explicit form of the modified Lorentz group:
\begin{equation}
\label{Lorentz}
\Lambda(x)^{\mu}_{\,\nu}=\left[e\right]^{\mu}_{\,a}\Lambda^{a}_{\,b}\left[e\right]^{\,b}_{\nu}
\end{equation}
obtaining a non-linear realization of this group, that preserves the form of the MDR and the homogeneity of degree $0$ of the perturbation functions.\\
The meaning of the formalism developed is that every particle lives in a section of the complete space-time, parameterized by its momentum. The \emph{tetrad} can be used to project vectors from a tangent space identified by the metric $g_{\mu\nu}(x,\,v)$ to a space with another metric $\overline{g}(y,\,w)_{\mu\nu}$ as summarized in the scheme below:
\[\begin{tikzcd}
         (TM,\,\eta_{ab},\,v) \arrow{d}{\left[e\right]} \arrow{rr}[swap]{\Lambda} && (TM,\,\eta_{ab},\,w)\arrow{d}[swap]{\left[\overline{e}\right]} \\
        (T_{x}M,\,g_{\mu\nu}(x,\,v)) \arrow{rr}[swap]{\left[\overline{e}\right]\circ\Lambda\circ\left[e^{-1}\right]} && (T_{x}M,\,\overline{g}_{\mu\nu}(y,\,w))
\end{tikzcd}\]
Now we introduce the modified connections of the constructed geometry, and starting from the definition of the \emph{Christoffel} one:
\begin{equation}
\label{Christoffel}
\Gamma_{\mu\nu}^{\,\alpha}=\frac{1}{2}g^{\alpha\beta}\left(\partial_{\mu}g_{\beta\nu}+\partial_{\nu}g_{\mu\beta}-\partial_{\beta}g_{\mu\nu}\right)
\end{equation}
using the explicit form of the metric (\ref{metric2}), it is simple to determine the connection components:
\begin{equation}
\label{connection1}
\begin{split}
&\Gamma_{\mu0}^{\,0}=\Gamma_{00}^{\,i}=\Gamma_{\mu\nu}^{\,i}=0\qquad\forall \mu\neq\nu\\
&\Gamma_{ii}^{\,0}=-\frac{1}{2}\partial_{0}f(p)\simeq0\\
&\Gamma_{0i}^{\,0}=\Gamma_{i0}^{\,0}=\frac{1}{2(1+f(p))}\partial_{0}f(p)\simeq0\\
&\Gamma_{ii}^{\,i}=\frac{1}{2(1+f(p))}\partial_{i}f(p)\simeq0\\
&\Gamma_{jj}^{\,i}=-\frac{1}{2(1+f(p))}\partial_{i}f(p)\simeq0\qquad\forall i\neq j\\
&\Gamma_{ij}^{\,i}=\Gamma_{ji}^{\,i}=\frac{1}{2(1+f(p))}\partial_{i}f(p)\simeq0\qquad\forall i\neq j
\end{split}
\end{equation}
where the latin indices vary inside the set $\{1,\,2,\,3\}$ and the greek ones inside $\{0,\,1,\,2,\,3\}$. For the not null terms, the approximation is possible because the interaction of a massive particle with the background is assumed tiny and the derivative $|\partial_{p}f(p)|\ll1$ is negligible because of the form of the perturbation functions (\ref{pert}). It is possible to introduce the local covariant derivative as:
\begin{equation}
\label{covder}
\nabla_{\mu}v^{\nu}=\partial_{\mu}v^{\nu}+\Gamma_{\mu\alpha}^{\,\nu}v^{\alpha}\simeq\partial_{\mu}v^{\nu} \, . 
\end{equation}
With this local covariant derivative we can compute the last connection that can determine the space-time, the \emph{Cartan} or \emph{spinorial} one, defined as:
\begin{equation}
\label{Cartan}
\omega_{\mu ab}=[e]^{\nu}_{\,a}\nabla_{\mu}[e]_{\nu b}\simeq[e]^{\nu}_{\,a}\partial_{\mu}[e]_{\nu b} \, . 
\end{equation}
Applying the first Cartan structural equation:
\begin{equation}
\label{struc}
de=e\wedge\omega
\end{equation}
to the external forms
\begin{equation}
\label{extform1}
\begin{split}
&e_{0}^{\,\mu}=dx^{\mu}\\
&e_{i}^{\,\mu}=\sqrt{1-f(p)} \, dx^{\mu}
\end{split}
\end{equation}
it is possible to show that, even for the spinorial connection, the not null elements are given by:
\begin{equation}
\label{extform2}
\frac{1}{2}\epsilon_{ijk}\omega^{ij}=\frac{1}{2}\frac{1}{1-f}\epsilon_{ijk}(\partial^{i}fdx^{j}-\partial^{j}fdx^{i}) \, . 
\end{equation}
Since they are proportional to derivatives of the perturbation functions, they are negligible, as in the previous case. Therefore, we can introduce the explicit form of the total covariant derivative of a tensor with a local index (greek) and a global one (latin):
\begin{equation}
\label{totalderivative}
D_{\mu}v^{\nu}_{\,a}=\partial_{\mu}v^{\nu}_{\,a}+\Gamma_{\mu\alpha}^{\,\nu}v^{\alpha}_{\,a}-\omega_{\mu\nu}^{\,a}v^{\nu}_{\,b}\simeq\partial_{\mu}v^{\nu}_{\,b}
\end{equation}
At this point, we can conclude that the introduction of a geometrized interaction, for massive particles with the ``quantized" background, identifies an asymptotically flat Finslerian structure.

\section{Standard Model extension}
\label{sec:4}
The introduced geometry modifications determine a change of the spinorial connection and, consequently, of the Dirac equation. In order to obtain the required changes, it is necessary to introduce the modified Dirac matrices, redefined in such a way to satisfy the Clifford Algebra relation:
\begin{equation}
\label{cliffalg}
\{\Gamma_{\mu},\Gamma_{\nu}\}=2g^{\mu\nu}=2[e]_{\mu}^{\,a}\eta_{ab}[e]_{\nu}^{\,b}
\end{equation}
that implies the following equality:
\begin{equation}
\label{gamma}
\Gamma^{\mu}=[e]^{\mu}_{\,a}\gamma^{a}
\end{equation}
From the previous requirement we obtain the explicit form of the modified Dirac matrices:
\begin{equation}
\label{modified-gamma}
\begin{split}
&\Gamma_{0}=\gamma_{0}\qquad \Gamma_{i}=\sqrt{1+f(p(x,\,\dot{x}))}\;\gamma_{i}\\
&\Gamma^{0}=\gamma^{0}\qquad \Gamma^{i}=\sqrt{1-f(p(x,\,\dot{x}))}\;\gamma^{i}
\end{split}
\end{equation}
and for the $\Gamma_5$ matrix:
\begin{equation}
\label{gamma5}
\begin{split}
&\Gamma_5=\frac{\epsilon^{\mu\nu\alpha\beta}}{4!}\Gamma_{\mu}\Gamma_{\nu}\Gamma_{\alpha}\Gamma_{\beta}=\frac{1}{\sqrt{\det{g}}}\Gamma_{0}\Gamma_{1}\Gamma_{2}\Gamma_{3}=\\
=&\frac{1}{\sqrt{\det{g}}}\sqrt{\det{g}}\,\gamma_{0}\gamma_{1}\gamma_{2}\gamma_{3}=\gamma_{5}\, , 
\end{split}
\end{equation}
where the total antisymmetric tensor $\epsilon_{\mu\nu\alpha\beta}$ for curved space-time has been used.\\
Finally the explicit form of the modified Dirac equation can be written as:
\begin{equation}
\label{Dirac-equation}
\left(i\Gamma^{\mu}\partial_{\mu}-m\right)\psi=0
\end{equation}
The modified equation that we derived  admits solutions which can be developed in plane waves, as in the standard case, and, resorting to the usual notation for spinors, we write them in the form:
\begin{equation}
\label{solspinor}
\begin{split}
&\psi^{+}(x)=u_{r}(p)e^{-ip_{\mu}x^{\mu}}\\
&\psi^{-}(x)=v_{r}(p)e^{ip_{\mu}x^{\mu}}
\end{split}
\end{equation}
The modified spinors can be easily computed in the usual way, considering the associated equation in momentum space, applied to the generic positive energy spinor:
\begin{equation}
\label{momeq}
(i\Gamma^{\mu}\partial_{\mu}-m)u_{r}(p)e^{-ip_{\mu}x^{\mu}}\Rightarrow(\slashed{p}-m)u_{r}(p)=0
\end{equation}
It is simple to derive the associated identity, for a spinor defined with null momentum $\overrightarrow{p}=0$:
\begin{equation}
\label{identity}
\begin{split}
&(\slashed{p}-m)(\slashed{p}+m)=(p^{\mu} p_{\mu})-m^2=0\;\Rightarrow \\
\Rightarrow&\;(\slashed{p}-m)(\slashed{p}+m)u_{r}(m,\,\overrightarrow{0})=0
\end{split}
\end{equation}
This equation implies that the generic spinor with momentum $\overrightarrow{p}$ can be obtained from the one with null momentum. From this statement the general form of a modified positive energy not normalized spinor immediately follows, using the standard representation of the Dirac matrices:
\begin{equation}
\label{modified-spinors}
\begin{split}
&(\Gamma^{\mu}p_{\mu}+m)\left(
                          \begin{array}{c}
                            \chi_{r} \\
                            0 \\
                          \end{array}
                        \right)
\Rightarrow\\
\Rightarrow&
\left(p^{0}\left(
             \begin{array}{cc}
               \mathbb{I} & 0 \\
               0 & \mathbb{-I} \\
             \end{array}
           \right)-
p^{i}\left(
       \begin{array}{cc}
         0 & -\sigma^{i} \\
         \sigma^{i} & 0 \\
       \end{array}
     \right)
\sqrt{1-f}\right)
\left(
  \begin{array}{c}
    \chi_{r} \\
    0 \\
  \end{array}
\right)+\\
+&m\left(
   \begin{array}{cc}
     \mathbb{I} & 0 \\
     0 & \mathbb{I} \\
   \end{array}
 \right)
 \left(
   \begin{array}{c}
     \chi_{r} \\
     0 \\
   \end{array}
 \right)
 =\left(
     \begin{array}{c}
       (E+m)\chi_{r} \\
       \overrightarrow{p}\overrightarrow{\sigma}\sqrt{1-f}\;\chi_{r} \\
     \end{array}
   \right)
\end{split}
\end{equation}
where the standard representation of the null momentum positive energy spinor has been used:
\begin{equation}
\label{standardspinor}
u_{r}(m,\,\overrightarrow{0})=\chi_{r}=\left(
                                         \begin{array}{c}
                                           1 \\
                                           0 \\
                                         \end{array}
                                       \right)
\end{equation}
The normalized form of this spinor (\ref{modified-spinors}) can be written as:
\begin{equation}
\begin{split}
\label{modified-spinors1}
&\left(
     \begin{array}{c}
       (E+m)\chi_{r} \\
       \overrightarrow{p}\overrightarrow{\sigma}\sqrt{1-f}\;\chi_{r} \\
     \end{array}
\right)\;\Rightarrow\\
\Rightarrow&\;u_{r}(m,\,\overrightarrow{p})=
\frac{1}{\sqrt{2m(E+m)}}
\left(
     \begin{array}{c}
       (E+m)\sqrt{1-f}\,\chi_{r} \\
       \overrightarrow{p}\overrightarrow{\sigma}\;\chi_{r} \\
     \end{array}
\right)
\end{split}
\end{equation}
All this derivation can be repeated with few changes to obtain an explicit form, for the negative energy spinors, analogous to that for the positive energy ones (\ref{modified-spinors}).\\
Starting from the new form of the relativistic Dirac
equation (\ref{Dirac-equation}), satisfied by the modified spinors $\psi$ defined in (\ref{modified-spinors1}), and from the modified gamma matrices  (\ref{modified-gamma}), one can define a modified current
\begin{equation}
\label{modified-current}
J^{\mu}=\bar{\psi}\Gamma^{\mu}\psi
\end{equation}
in such a way that there is a simplification between the corrections coming from the Lorentz violating coefficients present in the modified gamma matrices and spinors and, therefore, the modified current is defined in the normal tangent space $(T_{x}M,\eta_{\mu\nu})$. This brings to the following interaction term:
\begin{equation}
\label{QED-coupling}
J^{\mu}A_{\mu}=\bar{\psi}\Gamma^{\mu}\psi\eta_{\mu\nu} A^{\nu}\, \,
\end{equation}
describing the coupling with the electromagnetic field, which takes place in the tangent space $(T_{x}M,\eta_{\mu\nu})$.\\
The quark sector can be arranged, writing a modified effective Lagrangian of the form:
\begin{equation}
\label{quark-sector}
\frac{i}{2} \sum_{j}\bar{\psi}_{j}\Gamma^{\mu}\overleftrightarrow{D_{\mu}}\psi_{j}
\end{equation}
where $D_{\mu}$ represents the flat gauge covariant derivative of the SM. Even in this case, spinorial and Cartan connections
are negligible and we can globally conclude that our modified version of the Standard Model lives in an asymptotically flat space-time. Moreover, to preserve the $SU(3)$ internal symmetry of the strong interaction, the gauge fields of this theory are supposed to be Lorentz invariant as in the case of photons for QED. It is also remarkable that this kind of approach leads to the same results derived  in (\cite{Koste2}), where the modified Dirac matrices are defined as
$\Gamma^{\mu}=\gamma^{\mu}+c^{\mu\nu}\gamma_{\nu}$ and the authors adopt a particular choice of Lorentz-violating CPT even perturbation terms
of the form:
\begin{equation}
\label{57}
\frac{i}{2}\,c_{\mu\nu}\overline{\psi}\gamma^{\mu}\overleftrightarrow{D}^{\nu}\psi \, . 
\end{equation}

Following the previously introduced extensions of the SM, it is possible to proceed in the same way also for the weak part of the interaction and derive an effective theory representing the modified minimal extension of the usual Standard Model Lagrangian for electro-weak interactions. Using the modified Dirac matrices (\ref{modified-gamma}) and the modified spinors (\ref{modified-spinors1}), assuming temporary the neutrino masses equal to zero, as in Standard Model, it is simple to derive the explicit expression of the axial-vectorial current, which characterizes this interaction:
\begin{equation}
\label{avcurrent}
\bar{\psi}^{j}\Gamma^{\mu}(\mathbb{I}-\Gamma_{5})\tau^{i}\psi^{j}W^{i}_{\mu}=\bar{\psi}^{j}\Gamma^{\mu}(\mathbb{I}-\gamma_{5})\tau^{i}\psi^{j}W^{i}_{\mu}\, .
\end{equation}
where $\psi^{j}$ represents the doublet:
\begin{equation}
\label{57bis}
\psi^{j}=\left(
        \begin{array}{c}
          \nu^{j} \\
          l^{j} \\
        \end{array}
      \right)
\end{equation}
$l^{j}$ is the lepton associated to the $\nu^{j}$ neutrino, $W^{i}_{\mu}$ is the gauge boson mediating the charged electro-weak interaction and $\tau^{i}$ are the matrices representing the $SU(2)$ symmetry sector. These matrices are not modified, in order to preserve the symmetry of the interaction. Because $\Gamma_{5}=\gamma_{5}$ and the other modifications generated by LIV in spinors and Dirac matrices simplify, this current is defined in $(T_{x}M,\eta_{\mu\nu})$, as in case of QED, as already showed in eq.(\ref{modified-current}). 

Even in the more complete approach to electro-weak interactions, including quarks, the correction terms arrange in a similar way to what happens in the previous cases:  only the fermionic fields are modified, while the gauge fields are supposed to remain Lorentz invariant, in order to preserve the $SU(2)\times U(1)$ internal symmetry. Once more the corrections caused by the modified spinors simplifies with the ones generated by the modified Dirac matrices. From this immediately follows that LIV, as introduced in this work, only modifies the dynamics of massive particles, even neutrinos, without changing the interactions foreseen by the Standard Model. This means that the modified spinors maintain the same chirality as their standard counterpart and only the left handed particles  take an active part in the weak interaction.

\section{LIV and Neutrino oscillations in an Hamiltonian approach}
\label{sec:5}
Let's focus the attention on the central topic of this paper, that is the analysis of the eventual Lorentz violation effects  impact on
neutrino phenomenology, due to the modification of the flavor oscillation probabilities. This quantum phenomenon, which confirmed definitely that neutrino is a massive fermion, has been proved in a crystal-clear way both with natural neutrino sources (mainly solar~\cite{solar} and atmospheric~\cite{atmospheric}) and with artificial neutrinos (short~\cite{short-reactors} and long baseline~\cite{KamLAND} reactor antineutrinos, long-baseline~\cite{LBL-acceleratori} and, if one trusts the LSND~\cite{LSND} and MiniBOONE~\cite{MiniBoone} results, also short-baseline accelerator neutrino beams). The evidences of oscillation from disappearance experiments have been further reinforced in the last decade by appearance experiments, like the ones using the CNGS beam~\cite{OPERA} and like T2K~\cite{T2K} and No$\nu$A~\cite{NOVA} (which are collecting an increasing number of appearance signals of neutrinos with a flavor different from the production one).

As explained in the previous sections, the LIV perturbation introduced in our work can account just for tiny perturbative effects respect to the standard physics predictions. The presence of perturbative interaction terms, violating Lorentz invariance, determines a modification of the Hamiltonian $H$ that rules the evolution of neutrino wave function during its propagation, from the production point to the detector, according to Schroedinger equation: $i\partial_{t}|\psi\rangle=H|\psi\rangle$.\\
As we have seen, in a more general approach to LIV, the extended Standard Model Lagrangian can be written in the general form  \cite{Koste2}:
\begin{equation}
\label{L-SM-extended}
\mathcal{L} = \mathcal{L}_{0} + \mathcal{L}_{LIV}
\end{equation}
with
\begin{equation}
\label{L-LIV-general}
\mathcal{L}_{LIV}=-(a_{L})_{\mu}\overline{\psi}_{L}\gamma^{\mu}\overline{\psi}_{L}-(c_{L})_{\mu\nu}\overline{\psi}_{L}\gamma^{\mu}\partial^{\nu}\overline{\psi}_{L}
\end{equation}
The first term in eq.(\ref{L-LIV-general}), proportional to $(a_{L})$, violates CPT and, as a consequence, also Lorentz invariance, while the second contribution, proportional to $(c_{L})$, breaks ``only" Lorentz Invariance\footnote{It is well known from the work of Greenberg~\cite{Greenberg} that
LIV does not imply CPT violation. The opposite was, instead, declared to be true in the same work~\cite{Greenberg}. However, the fact that CPT violation
automatically brings to LIV was confuted in~\cite{Dolgov-et-al} (where a counterexample was found, by considering a nonlocal model) and the argument has been widely debated in literature~\cite{Discussion-LIV-CPT}.}.
Consequently, it is possible to build the effective LIV Hamiltonian as the following sum:
\begin{equation}
\label{effective-LIV-Hamiltonian}
H_{eff} = H_{0} + H_{LIV}
\end{equation}
where $H_{0}$ denotes the usual Hamiltonian conserving Lorentz invariance and $H_{LIV}$ indicates the corrections introduced by the tiny LIV violating terms of (\ref{L-LIV-general}). Using a perturbative approach and neglecting the part of $H_{0}$ that (for a fixed momentum neutrino beam) contributes identically to all the three mass eigenvalues and, therefore, do not influence the oscillation probability, the remaining part of the extended Hamiltonian can be written as:
\begin{equation}
H = \frac{1}{2E}\left(M^2+2(a_{L})_{\mu}p^{\mu}+2(c_{L})_{\mu\nu}p^{\mu}p^{\nu}\right)
\end{equation}
where $M^2$ is a $3\times3$ matrix, that in the mass eigenvalues basis assumes the form:
\begin{equation}
\label{61}
\left(\begin{array}{ccc}
         m_{1}^{2} & 0 & 0 \\
         0 & m_{2}^{2} & 0 \\
         0 & 0 & m_{3}^{2} \\
      \end{array}\right)
\end{equation}
\\ 
Resorting to the quantum mechanic perturbation theory, the new eigenstates become:
\begin{equation}
\label{64}
|\widetilde{\nu}_{i}\rangle=|\nu_{i}\rangle+\sum_{i\neq j}\frac{\langle\nu_{j}|H_{LIV}|\nu_{i}\rangle}{E_{i}-E_{j}}|\nu_{j}\rangle
\end{equation}
It is possible to define the perturbed time evolution operator, as in the work~\cite{Diaz:2009qk}:
\begin{equation}
\label{65}
\begin{split}
&S(t)=\left(e^{-(iH_{0}+H_{LIV})t} e^{iH_{0}t}\right)e^{-iH_{0}t}=\\
=&\left(e^{-i(H_{0}+H_{LIV})t} e^{iH_{0}t}\right)S^{0}(t)
\end{split}
\end{equation}
and to evaluate the oscillation probability as:
\begin{equation}
\begin{split}
&P(\nu_{\alpha}\rightarrow\nu_{\beta})=|\langle\beta(t)|\alpha(0)\rangle|^2=\\
&\Biggr|\sum_{n}[\langle\beta(t)|\left(|n_{0}\rangle\langle n_{0}|+\sum_{j\neq n}\frac{\langle j_{0}|H_{LIV}|n_{0}\rangle}{E^{0}_{n}-E^{0}_{j}}|j_{0}\rangle\langle j_{0}|\right)|\alpha(0)\rangle+\\
&\ldots]\Biggr|^2 = P^{0}(\nu_{\alpha}\rightarrow\nu_{\beta})+P^{1}(\nu_{\alpha}\rightarrow\nu_{\beta})+\ldots\\
\label{oscillation-probability}
\end{split}
\end{equation}

In eq.(\ref{oscillation-probability}) $P^{0}(\nu_{\alpha}\rightarrow\nu_{\beta})$ represents the usual foreseen oscillation probability and the other term is:
\begin{equation}
\label{P1-osc}
\begin{split}
&P^{1}(\nu_{\alpha}\rightarrow\nu_{\beta})=\\
=&\sum_{ij}\sum_{\rho\sigma}
2L\,\mathfrak{Re}\left(\left(S^{0}_{\alpha\beta}\right)^{*} U_{\alpha i}U_{\rho i}^{*}H^{LIV}_{\rho\sigma}U_{\sigma j}U_{\beta j}^{*}\tau_{ij}\right)
\end{split}
\end{equation}
with:
\begin{equation}
\label{unitary}
U_{\alpha i}=\langle\alpha|i\rangle
\end{equation}
where $|\alpha\rangle$ denotes a flavor eigenstate and, instead, $|j\rangle$ represents a $H_{0}$ one, that is a mass eigenstate. Moreover in (\ref{P1-osc}):
\begin{equation}
\label{68}
\tau_{ij}=\left\{
            \begin{array}{ll}
              (-i)e^{-iE_{i}t}\qquad i=j\\
              \frac{e^{-iE_{i}t}-e^{-iE_{j}t}}{E_{i}-E_{j}}\qquad i\neq j
            \end{array}
          \right.
\end{equation}
with the constrains on the Hamiltonian matrix:
\begin{equation}
\label{69}
\left\{
  \begin{array}{ll}
    H^{LIV}_{\alpha\beta}=\left(H^{LIV}_{\beta\alpha}\right)^{*}\qquad\alpha\neq\beta \\
    H^{LIV}_{\alpha\alpha}\in\mathbb{R}
  \end{array}
\right.
\end{equation}
Hence, also the flavor transition probability can be expressed, as expected, in terms of a perturbative expansion. In case of a general treatment of $H_{LIV}$, assuming a direction depending perturbation, it would be necessary to specify a privileged frame of reference when reporting this kind of results.\\

\section{LIV and Neutrino oscillations in our model}
\label{sec:6}
Another equivalent way to introduce LIV in neutrino oscillations consists in using directly the Modified Dispersion Relations, as we do in geometrizing the neutrino interactions with the background. In this work we assume that the MDRs of neutrinos are spherically symmetric (as already done in literature in  the so called ``\emph{fried chicken models}"~\cite{chicken}). Until now there are no experimental evidences against this assumption. In this way the eq.(\ref{mdr3}), expressing MDRs, reduces to the form:
\begin{equation}
\label{mdr5}
E^2=|\overrightarrow{p}|^2\left(1-f\left(\frac{|\overrightarrow{p}|}{E}\right)\right)+m^2\, .
\end{equation}
Furthermore, using the homogeneity of degree $0$ of the perturbation function $f$, we have shown that the MDR is originated by a metric in the momentum space and this guarantees the validity of Hamiltonian dynamics. As just underlined, the propagation in vacuum of an ultra-relativistic particle, such as a neutrino, is governed by the Schr\"odinger equation, whose solutions are in the form of generic plane waves:
\begin{equation}
\label{planewave}
e^{i(p_{\mu}x^{\mu})}=e^{i(Et-\overrightarrow{p}\cdot\overrightarrow{x})}=e^{i\phi}
\end{equation}

The effects of the modified metric do not appear, because the contraction is between a covariant and a controvariant vector, so the correction terms simplify. To give the explicit form of the solution, we start from the MDR (\ref{mdr5}) 
and, using the approximation of ultrarelativistic particle $|\overrightarrow{p}|\simeq E$, we obtain:
\begin{equation}
\label{neutrinoenergy}
\begin{split}
 |\overrightarrow{p}|=&\, \sqrt{|\overrightarrow{p}|^2\left(1-f\left(\frac{|\overrightarrow{p}|}{E}\right)\right)+m^2} \simeq\\\,
\simeq
&
\, \, \, E\left(1-\frac{1}{2}f\left(\frac{|\overrightarrow{p}|}{E}\right)\right)+\frac{m^2}{2E}
\end{split}
\end{equation}
In this way it is possible to evaluate the phase $\phi$ of the plane wave of eq.(\ref{planewave}) for a given mass eigenstate, resorting to the natural measure units for which $t=L$:
\begin{equation}
\label{phase}
\phi=Et-EL+\frac{f}{2}EL-\frac{m^2}{2E}L=
\left(f E -\frac{m^2}{E} \right) \frac{L}{2}\, .
\end{equation}
Hence the phase difference of two mass eigenstates for neutrinos with the same energy $E$ can be written as:
\begin{equation}
\label{diffphase}
\begin{split}
\Delta\phi_{kj} = & \, \phi_{j}-\phi_{k} =
\frac{(f_{j}-f_{k})}{2}EL-\left(\frac{m_{j}^{2}}{2E}-\frac{m_{k}^{2}}{2E}\right)L=\\
=&\left(\frac{\Delta m_{kj}^{2}}{2E}-\frac{\delta f_{kj}}{2}E\right)L
\end{split}
\end{equation}
The oscillation probability depends on the phase differences $\Delta\phi_{kj}$, in addition to the usual $3\times3$ unitary matrix PMNS, and the transition probability from a flavor $|\alpha\rangle$ to a flavor $|\beta\rangle$ (in the most general case, including even the CP violating phase in the mixing matrix) can be written in the usual form:
\begin{equation}
\label{probability}
\begin{split}
P(\nu_{\alpha}\rightarrow\nu_{\beta}) = &\delta_{\alpha\beta}-4\sum_{i>j}\mathfrak{Re}\left(U_{\alpha i}U_{\beta i}^{*}U_{\alpha j}^{*}U_{\beta j}\sin^2(\Delta\phi_{ij})\right)+\\
+&2\sum_{i>j}\mathfrak{Im}\left(U_{\alpha i}U_{\beta i}^{*}U_{\alpha j}^{*}U_{\beta j}\sin^2(\Delta\phi_{ij})\right)
\end{split}
\end{equation}
We are in presence of a modified oscillation probability, due to the appearance in the phase differences (defined in eq.(\ref{diffphase}) of the LIV violating correction term proportional to $\delta f_{kj} = f_k - f_j$. This term is different from zero only if the coefficients $f_i$ ruling the LIV violations are not equal for all the three
mass eigenstates; otherwise the expression of eq.(\ref{probability}) reproduces the usual  three flavor oscillation probability one gets in absence of Lorentz invariance violation.

It is essential to notice that our model, considering CPT even LIV terms (introduced starting from MDR),  deals only with oscillation effects caused by the difference of LIV perturbations to different mass eigenstates (\cite{Liberati2}). The fundamental assumption, that represents a reasonable physical hypothesis, is that every mass state presents a personal maximum attainable velocity, because does not interact in the same way of the others with the background.
It is even important to underline that the form of Lorentz invariance violation introduced in our model couldn't explain the neutrino oscillation without the introduction of masses. In fact the perturbative mass term, introduced by our LIV theory, is proportional to the energy of the particle, and this would be in contrast with the evidences of neutrino oscillations, if were not present a dominant mass term that does not present such a dependence. Neutrino oscillations are well described by models with the phase depending only on squared masses differences divided by the energy:
\begin{equation}
\label{standardphase}
\Delta\phi_{jk}=\left(\frac{m_{j}^2}{2E}-\frac{m_{k}^2}{2E}\right) L=\frac{\Delta m_{jk}^2}{2E}L
\end{equation}
and LIV effects, of the type here introduced, could only appear at high energies as tiny perturbations (\ref{diffphase}). This theory, therefore, can only account for relatively  little deviations from what is considered ``standard physics'' and, in neutrino oscillation sector, could generate only little effects, at the highest observable energies. Nevertheless, experimentally these effects would be very interesting, because they could be observable, representing modifications of the classical theory predictions for the spectrum.

Other LIV theories can even explain oscillations without resorting to the classical concept of mass~\cite{Arias};  in fact they usually insert terms in the Standard Model Lagrangian that introduce masses for neutrinos as caused by the interaction with background fields, as in~\cite{Kostelecky-mancante}, where the modified Dirac equation can be written using the modified Dirac matrices:
\begin{equation}
\label{k1}
\begin{split}
\Gamma_{AB}^{\,\mu}=&\gamma^{\mu}\delta_{AB}+c_{AB}^{\mu\nu}\gamma_{\nu}+d_{AB}^{\mu\nu}\gamma_{5}\gamma_{\nu}+\\
+&e^{\,\mu}_{AB}+if^{\mu}_{AB}\gamma_{5}+\frac{1}{2}g_{AB}^{\mu\nu\tau}\sigma_{\nu\tau}
\end{split}
\end{equation}
and the modified mass matrix:
\begin{equation}
\begin{split}
\label{k2}
M_{AB}=&m_{AB}+im_{5AB}\gamma_{5}+a^{\,\mu}_{AB}\gamma_{\mu}+\\
+&b^{\,\mu}_{AB}\gamma_{5}\gamma_{\mu}+\frac{1}{2}H_{AB}^{\mu\nu}\sigma_{\mu\nu}
\end{split}
\end{equation}
In the previous equations $m$ and $m_{5}$ are Lorentz and CPT conserving masses. The CPT conserving Lorentz violating terms are: $c,\,d,\,H$, while $a,\,b,\,e,\,f,\,g$ are CPT violating.

In this case, the LIV introduced mass terms would constitute a theoretical justification for the oscillations, but this kind of LIV introduced masses (differently from the CPT even LIV corrections present in our model and also in~\cite{Koste2}) would not spoil the general dependence of the oscillation probabilities on the neutrino energy and, therefore, it would not modify the ``standard'' oscillation pattern
with the introduction of new effects that could be experimentally used to confirm or not the validity of LIV hypothesis.
\section{Phenomenological analysis of the LIV effects on neutrino oscillations}
\label{sec:7}

It is well known that neutrino physics, thanks to its very rich and various set of experiments (covering a wide spectrum of energies and baselines), is an ideal playground to search for deviations from Lorentz invariance~\cite{neutrino-and-LIV}.

In order to evaluate the impact on neutrino phenomenology of the possible sources of Lorentz invariance violations
studied in our model, we compared the three oscillation probabilities ruling the possible neutrino oscillations ($P_{\nu_e,\nu_{\mu}}$, $P_{\nu_e,\nu_{\tau}}$ e $P_{\nu_{\mu},\nu_{\tau}}$)
evaluated (by means of equations (\ref{probability}) and (\ref{diffphase})) in presence of LIV, with
the standard oscillation probabilities one gets if Lorentz invariance is satisfied.

Differently from other previous studies 
 that adopted the two flavor oscillation approximation, our analysis has been pursued in the realistic three flavor scenario.
The values of the $\Delta m^2_{ij}$ and of the various PMNS matrix elements ($U_{\alpha,i}$) entering the calculation have been taken by the most recent global fits including all the different neutrino experiments~\cite{Lisi-et-al,fit-altri}. We assumed for simplicity  the value $\delta = 0$ for the Dirac CP violation phase, because in our case we are not interested in the study of CP violating effects (which would not spoil our results), but the analysis could be easily modified to introduce also this effect.

The outcome of our study is reported in the following series of figures,
where we draw the different oscillation probabilities $P_{\nu_{\alpha}, {\nu_{\beta}}}$, in absence and in presence of Lorentz violating terms, evaluated, for fixed values of the neutrino energy,
as a function of the baseline L (that is the distance between the neutrino production and detection points).
The first two graphs report the probabilities for a muonic neutrino to oscillate, respectively, into an electronic and a tauonic one.
The probability $P_{\nu_{\mu}, \nu_{\tau}}$ is the most relevant one for the study of atmospheric neutrinos and it is important also for long-baseline
accelerator neutrino experiments;  the knowledge of $P_{\nu_{\mu}, \nu_{e}}$
over a wide range of L (from 1 up to $10^5-10^6$  km) covers the regions of interest both for short- and long-baseline accelerator experiments and also for reactor antineutrino experiments (because $P_{\bar{\nu}_{e}, {\bar{\nu}}_{\mu}} = P_{\nu_{\mu}, \nu_{e}}$ under the assumption of CPT invariance).
The remaining oscillation probability $P_{\nu_{e}, \nu_{\tau}}$ is shown, instead, in fig..\ref{etau1gev}.
The value (E = 1 GeV) considered for the energy in this series of 3 figures has been chosen having in mind the order of magnitude of the characteristic energies relevant for the oscillation studies both for atmospheric and for long-baseline accelerator neutrino experiments.
\begin{figure}[h]
\includegraphics[width=80mm]{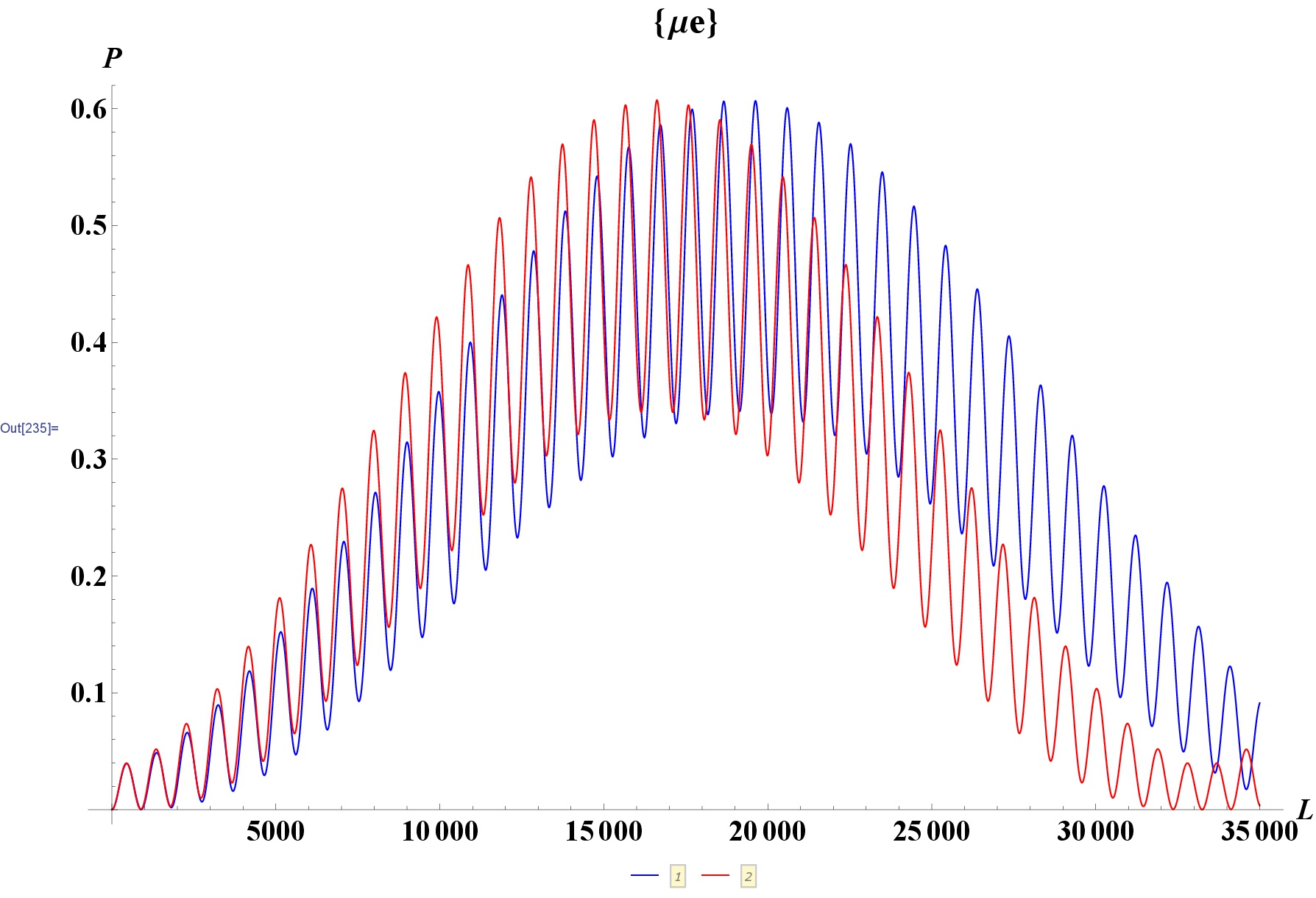}
\caption{Comparison of the oscillation probability from $\nu_{\mu}$ to $\nu_e$, computed (as a function of the baseline L) for neutrino energy
${\rm E=1 \, GeV}$, in the ``standard theory" (red curve) and in presence of LIV (blue curve), for LIV parameters $\delta f_{32} = \delta f_{21} = 1 \times 10^{-23}$.}
\label{mue1gev}
\end{figure}
\begin{figure}[h]
\includegraphics[width=80mm]{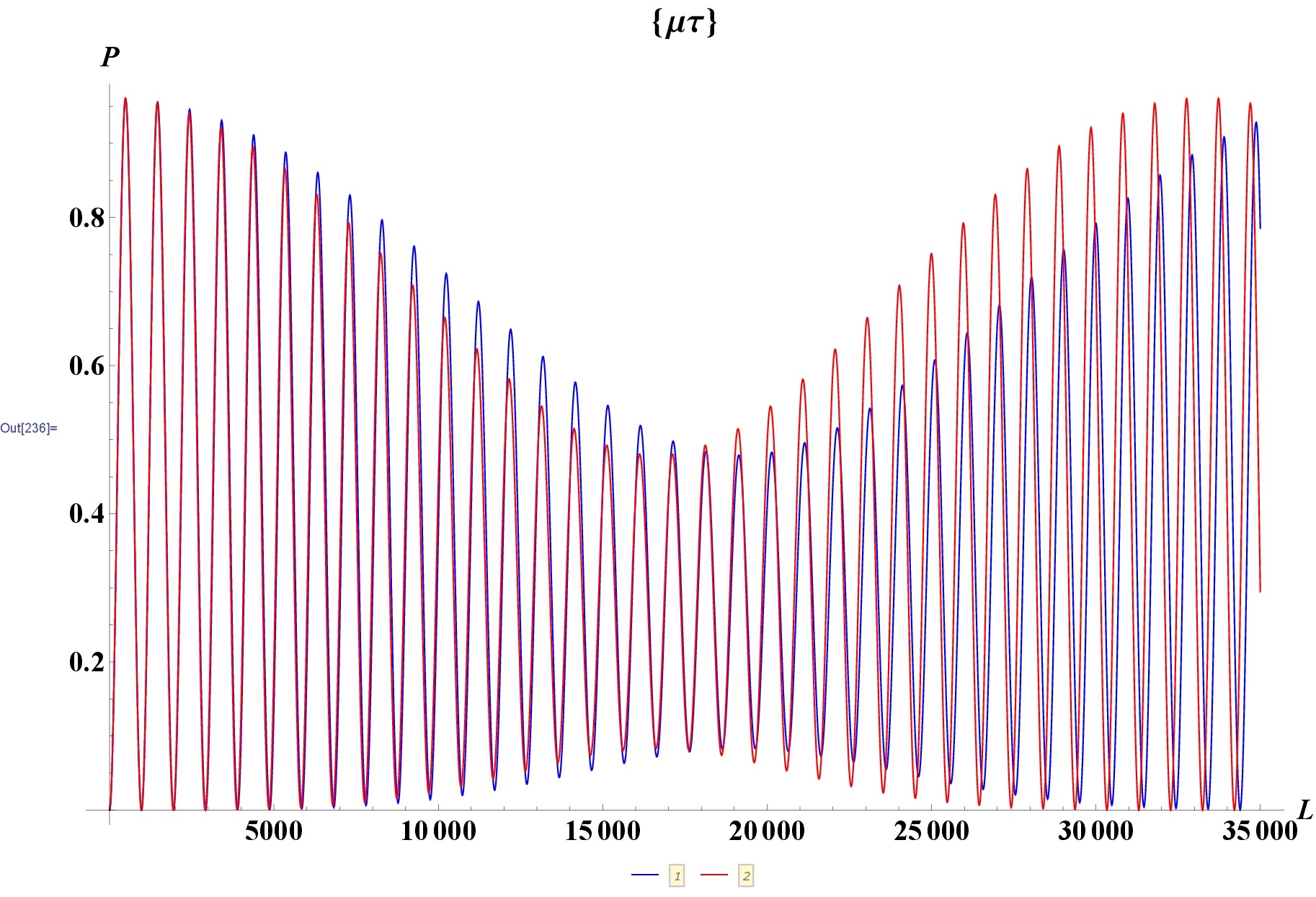}
\caption{Same analysis of  fig.\ref{mue1gev}, but for the oscillation probability from $\nu_{\mu}$ to
$\nu_{\tau}$.}
\label{mutau1gev}
\end{figure}
\begin{figure}[h]
\includegraphics[width=80mm]{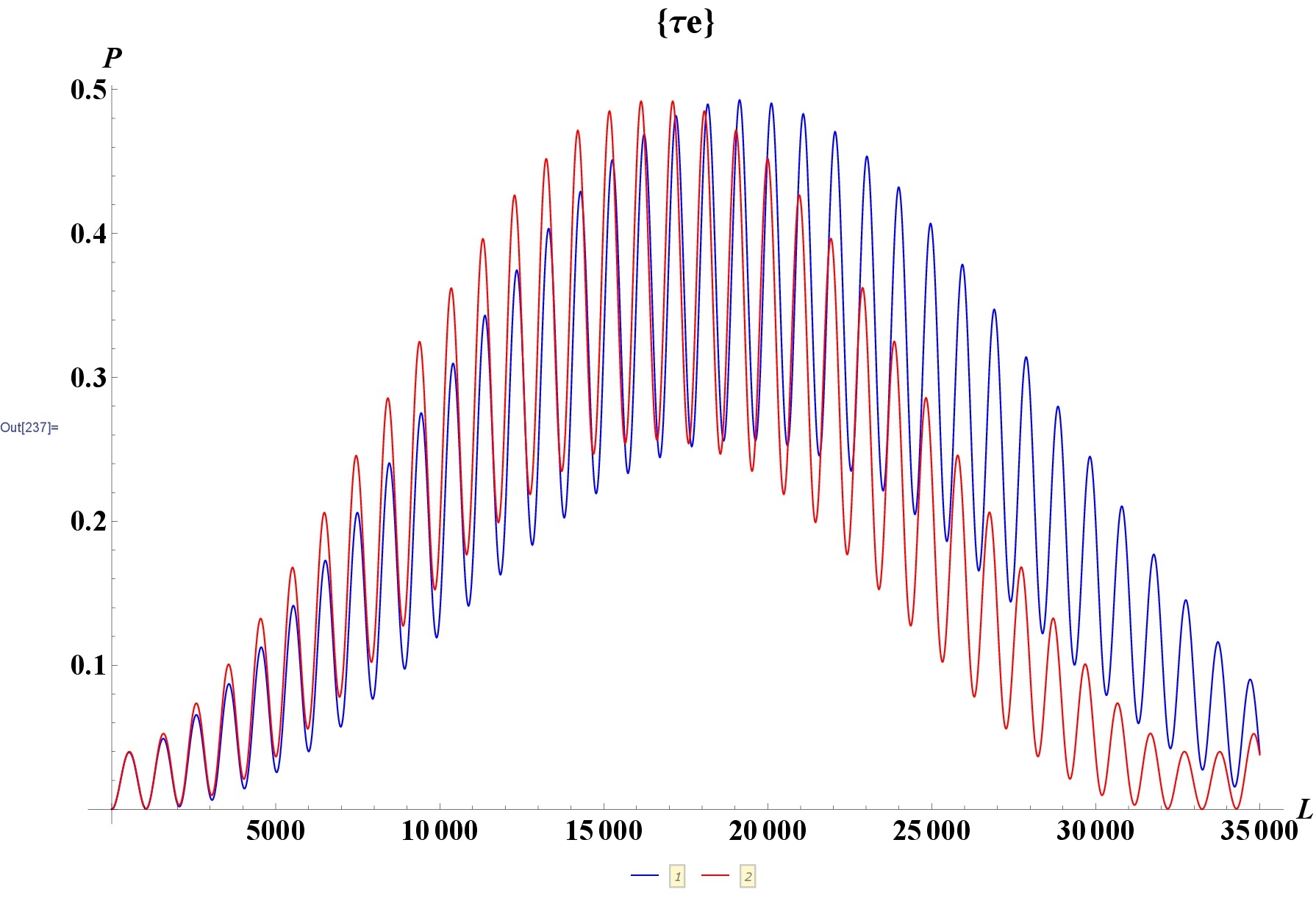}
\caption{Same analysis of  fig.\ref{mue1gev}, but for the oscillation probability from
$\nu_{e}$ to $\nu_{\tau}$.}
\label{etau1gev}
\end{figure}

The order of magnitude of the oscillation probability corrections induced by the Lorentz invariance violations is determined by the values chosen for the three parameters $f_k$ and, consequently, for their differences
$\delta f_{kj}$, as shown in eq.(\ref{diffphase}). We assumed for simplicity that the 3 parameters $f_k$ are of same order of magnitude and that they are ordered in a ``natural" way, with the highest LIV parameter correction associated to the highest mass eigenvalue (that is: $f_1 < f_2 < f_3$ and
$\delta f_{32} \simeq \delta f_{21}$).
In figs.\ref{mue1gev}-\ref{etau1gev}
we adopted the values $\delta f_{32} = \delta f_{21}= 1 \times 10^{-23}$, which are of the same order of magnitude of the limits derived for LIV violation
in the phenomenological studies one could find in literature up to 2015~\cite{altri-fenomenologici}, or even more conservative
than these limits.
As one can see clearly from figs.\ref{mue1gev}-\ref{etau1gev}, for
$\delta f_{ki} = 1 \times 10^{-23}$ the presence of LIV would modify in a visible way the oscillation
probabilities patterns.

However, recently the SuperKamiokande collaboration performed a test of Lorentz
invariance by analyzing atmospheric neutrino data and derived more stringent constraints on the possible values of the coefficients for Lorentz invariant violating corrections to the Hamiltonian~\cite{SK-test-LIV}. In particular, limits of the order of $10^{-26}-10^{-27}$  were derived for the coefficient of the isotropic CPT even term, that introduces corrections to the oscillation probabilities  proportional to $L \times E$ and would correspond to the kind of Lorentz invariance violation of our model. As a matter of fact, the comparison between our model and the
Hamiltonian assumed as a reference for the SuperK analysis is not so immediate, because in that Hamiltonian are present also other kinds of LIV violating corrections and in particular CPT odd terms (introducing corrections to $P_{\nu_{\alpha},\nu_{\beta}}$ not proportional to the neutrino energy) of the order of $10^{-23}$.

\begin{figure}[h]
\includegraphics[width=70mm]{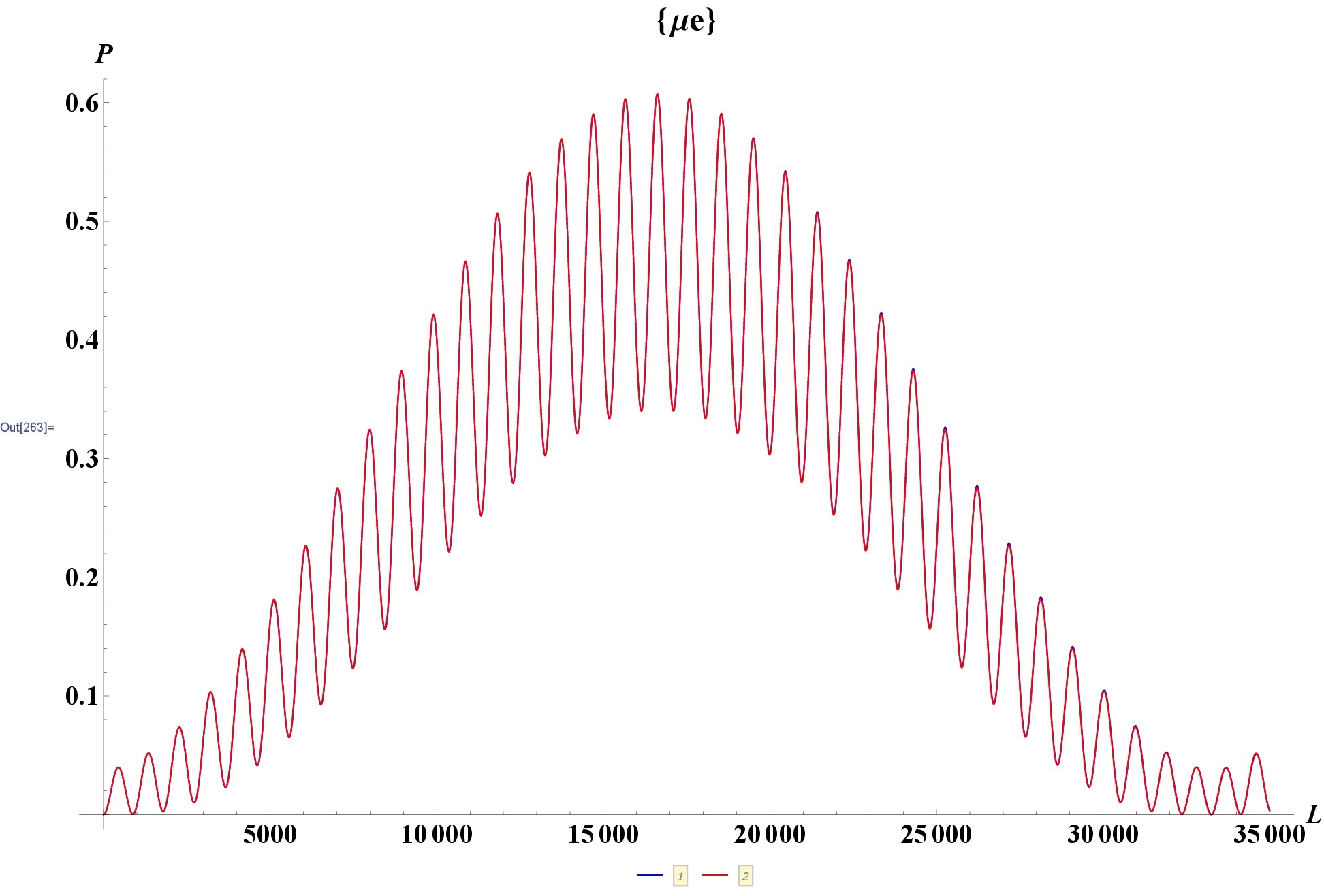}
\caption{ Same analysis of  fig.\ref{mue1gev}, but for LIV parameters $\delta f_{kj}$ of the order of $10^{-25}$.}
\label{mue1gevpiccolo}
\end{figure}

The fig.\ref{mue1gevpiccolo} reports the comparison of the $\nu_{\mu}-\nu_e$ oscillation probabilities with and without LIV for
values of our parameters
$\delta f_{kj} = 10^{-25}$. In this case the two curves are practically superimposed and the situation is essentially the same also for
$P_{\nu_{\mu},\nu_{\tau}}$ and for
$P_{\nu_{e},\nu_{\tau}}$.
The effects of LIV corrections are not anymore visible and the percentage variations of $P_{\nu_{\alpha},\nu_{\beta}}$ are lower than $1 \%$ essentially
in all the regions in which P is significantly different from zero.
%

Therefore, for values of the $\delta f_{kj}$ coefficients of the same order derived by SuperKamiokande
for the CPT even isotropic LIV corrections ($\delta f_{kj} \simeq 10^{.-26}-10^{-27}$), the LIV effects on the oscillation probabilities are observable only for higher neutrino energies.

\begin{figure}[h]
\includegraphics[height=70mm]{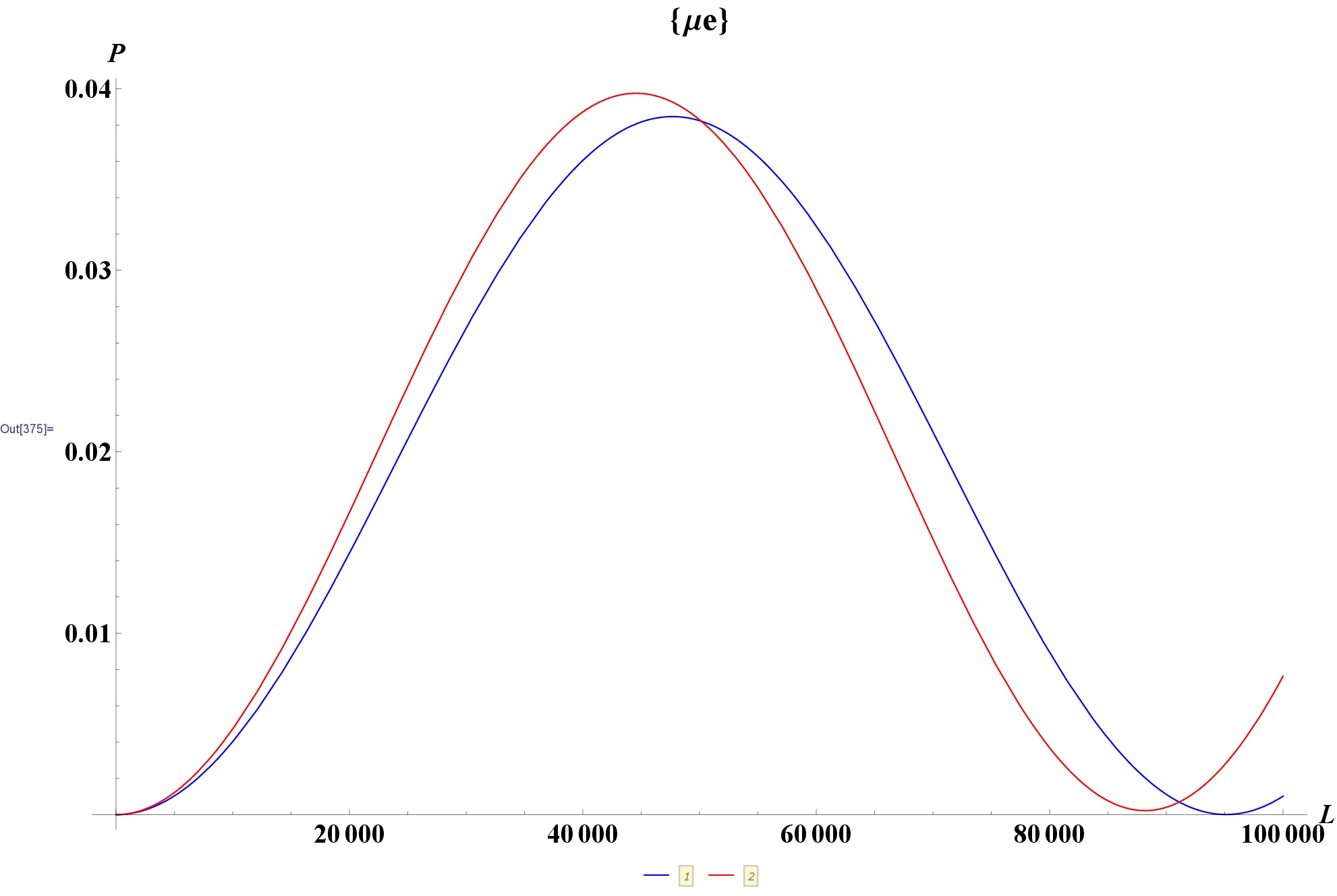}
\caption{ Same analysis of  fig.\ref{mue1gev}, but for LIV parameters $\delta f_{32} = \delta f_{21} = 4.5 \times 10^{-27}$   and for neutrino energy {\rm E = 100 GeV}.}
\label{mue100gev}
\end{figure}

\begin{figure}[h]
\includegraphics[height=60mm]{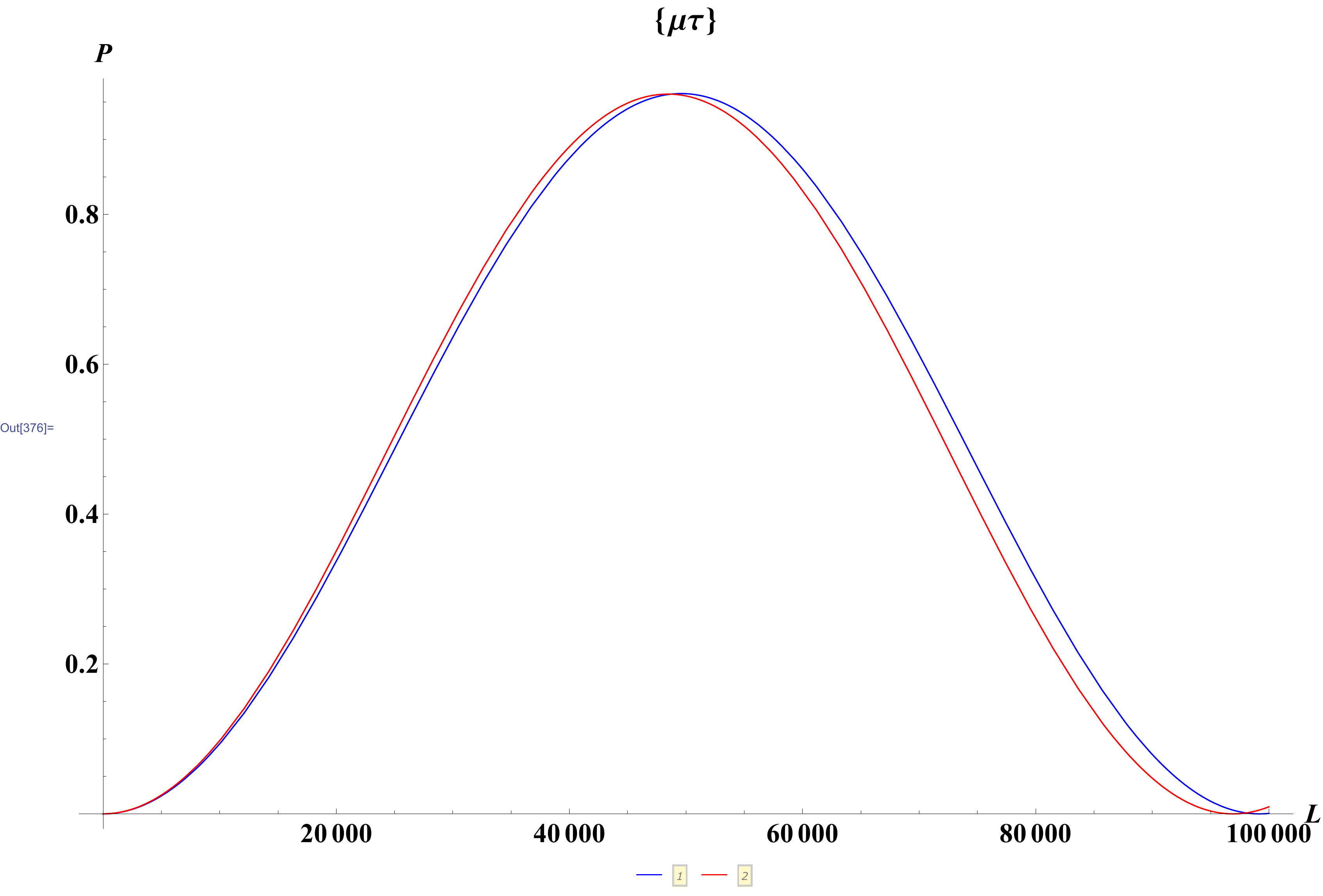}
\caption{Same of  fig.\ref{mue100gev}, but for the oscillation probability
$P_{\nu_{\mu},\nu_{\tau}}$.}
\label{mutau100gev}
\end{figure}

\begin{figure}[h]
\includegraphics[width=80mm]{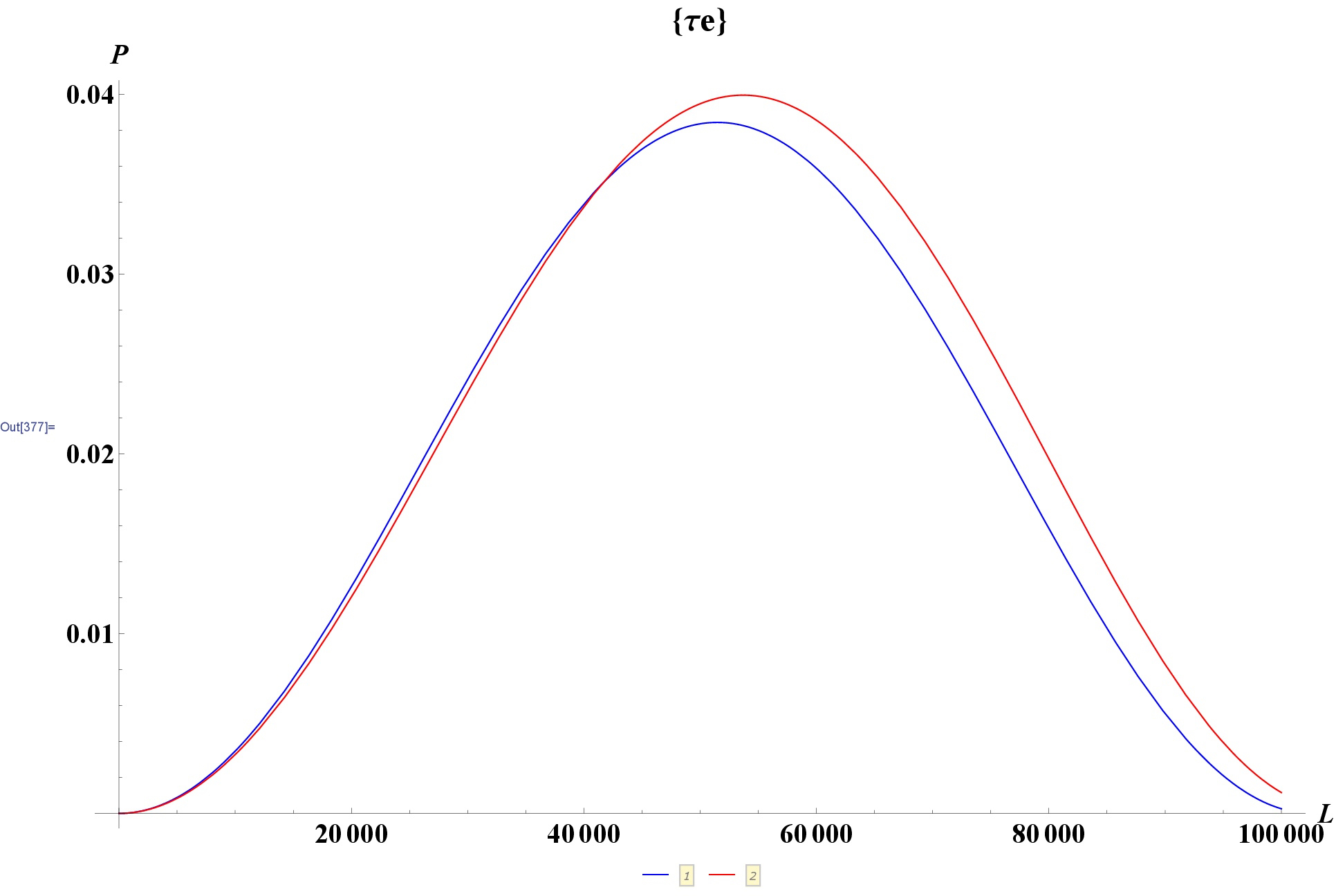}
\caption{Same of  fig.\ref{mue100gev}, but for $P_{\nu_{e}, \nu_{\tau}}$.}
\label{etau100gev}
\end{figure}

\begin{figure}
\includegraphics[width=80mm]{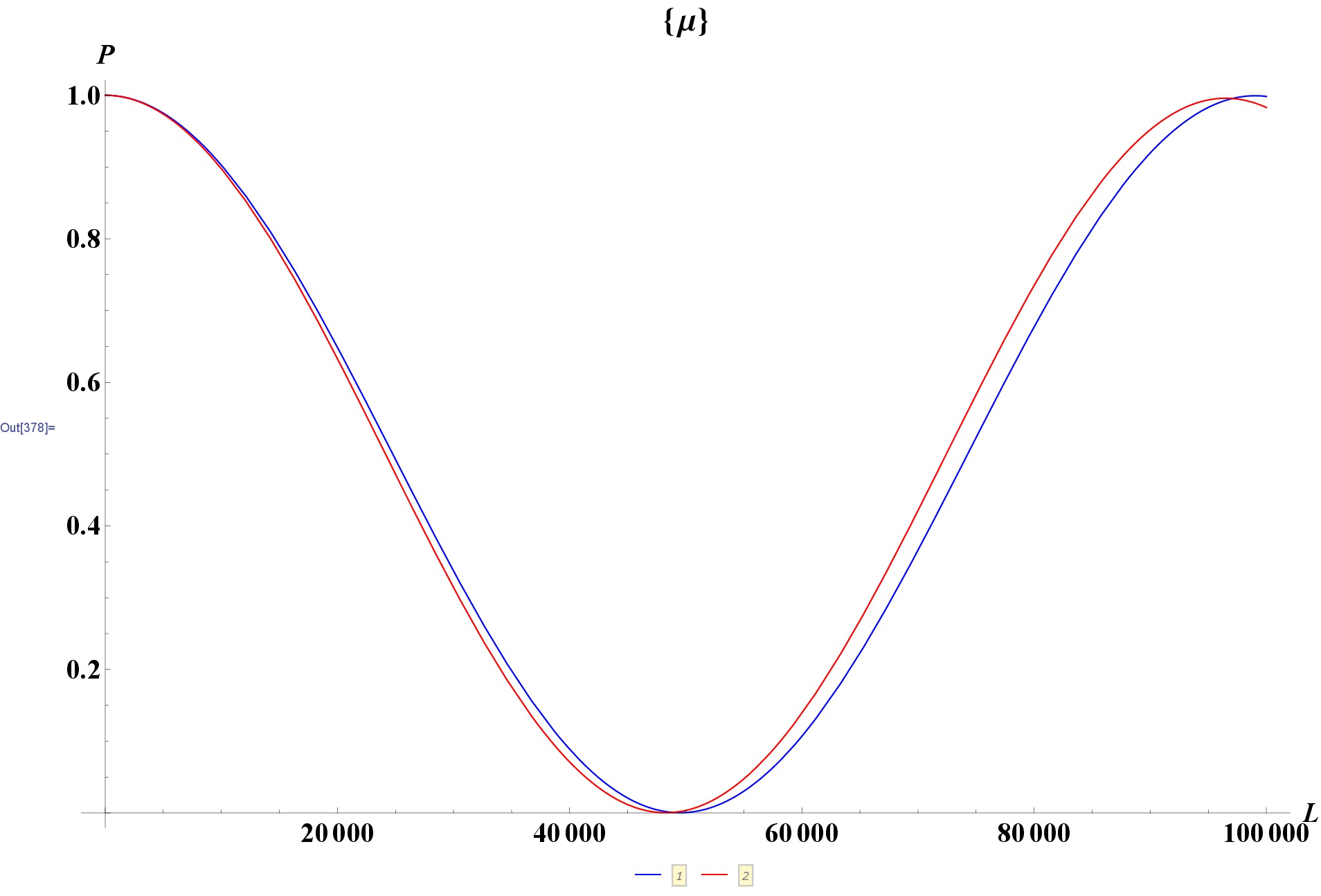}
\caption{Total survival probability for muonic neutrino, evaluated for the same conditions of fig.\ref{mue100gev}}
\label{mumu100gev}
\end{figure}

\begin{figure}[h]
\includegraphics[width=85mm]{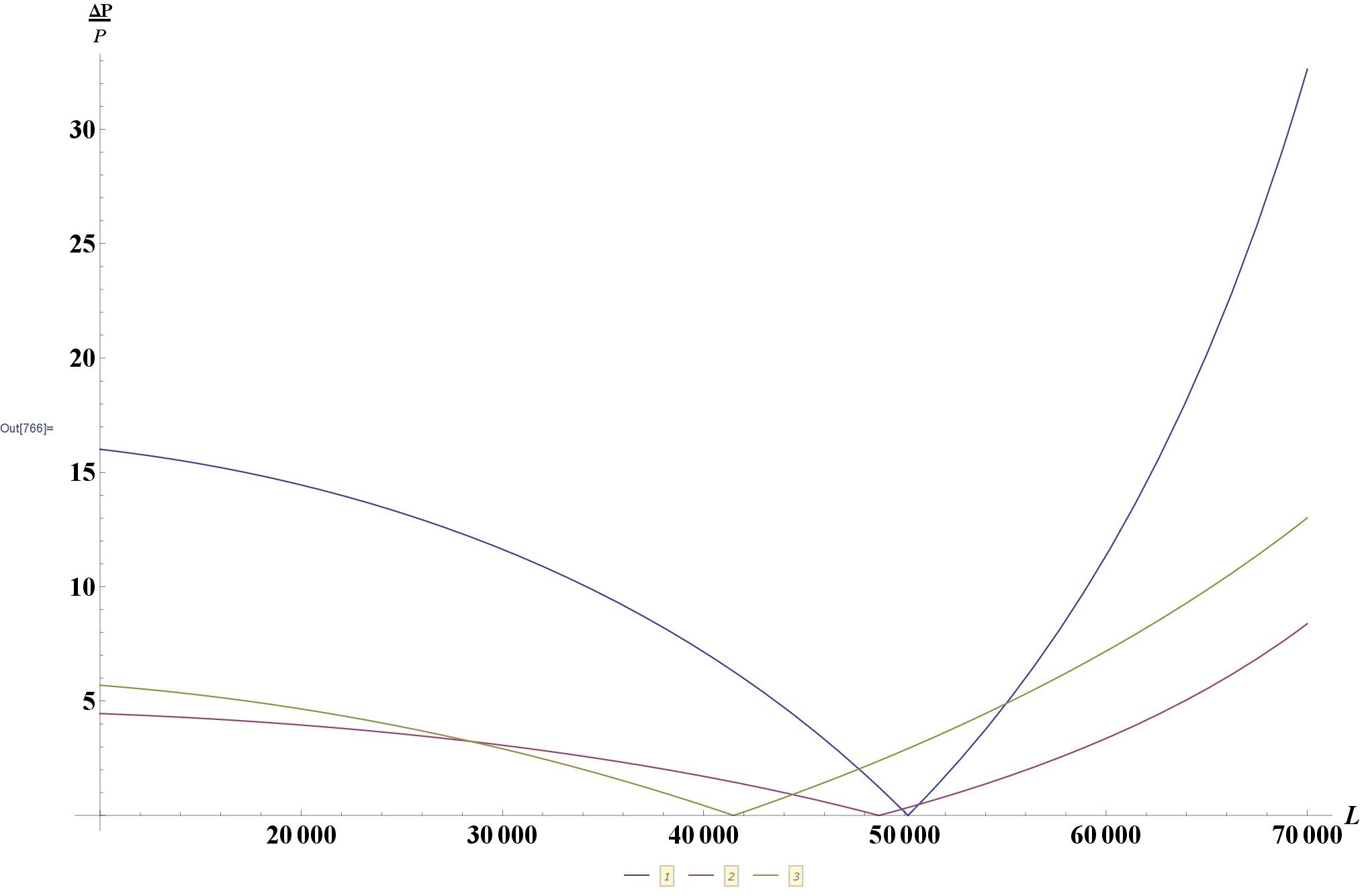}
\caption{Percentage variations induced in the neutrino oscillation probabilities by the LIV corrections.
On the vertical axis we report,  as a function of the baseline L, the percentage differences between the oscillation probabilities for
a 100 GeV neutrino, in presence and in absence of LIV, normalized with respect to their average value. The 3 different curves
correspond to the percentage differences for the 3 oscillation probabilities: $P_{\nu_e,\nu_{\mu}}$ (blue),
$P_{\nu_{\mu},\nu_{\tau}}$ (violet) and $P_{\nu_{e},\nu_{\tau}}$ (green curve).}
\label{differenze}
\end{figure}

In the figs.\ref{mue100gev}-\ref{mumu100gev} we report the results we obtained for the three oscillation probabilities and for the total $\nu_{\mu}$ survival probability ($1-P_{\nu_{\mu}.\nu_e}-P_{\nu_{\mu},\nu_{\tau} }$) in the case of a neutrino of 100 GeV, an energy that is studied, for instance, for atmospheric neutrinos by SuperKamiokande and by the neutrino telescopes. In these graphs we assumed
$\delta f_{32} = \delta f_{21} = 4.5 \times 10^{-27}$, that is of the same order of magnitude derived for the corresponding parameter
by SuperKamiokande. For these values of the coefficients the LIV effects are visible and they induce variations of the oscillations and survival probabilities of at least a few percent for most values of L, as one can see directly by fig.\ref{differenze}. In this figure we represented simultaneously (for all the 3 probabilities $P_{\nu_{\alpha},\nu_{\beta}}$)
the percentage variations due to LIV corrections
(computed as $2\, \frac{P_{\small LIV}-P_{\small NOLIV}}{P_{\small LIV}+P_{\small NOLIV}} \times 100$), evaluated over a restricted range of values for the baseline, $ {\rm 10000 \, km < L < 70000 \, km}$, for which the oscillation probabilities are not too low. For most of the values considered for the baseline, the LIV induced percentage variations are higher than $5-10 \%$ for $P_{\nu_{\mu},\nu_e}$ and above $2-3 \%$ for the two other oscillation probabilities. In the range considered, the LIV corrections become particularly significant for ${\rm L  > 60000 \, km}$ (more than $15\, \%$ for $P_{\nu_{\mu},\nu_e})$~\footnote{A word of caution must be spent about the interpretation of these percentage variations, that must be evaluated considering also the absolute value of the oscillation probability, used to ``normalize'' these variations. For some values of L,  higher percentage variations sometimes are mainly due to the fact that the corresponding absolute value of $P_{\nu_{\alpha},\nu_{\beta}}$ is extremely small.}.


\begin{figure}[h]
\includegraphics[width=90mm]{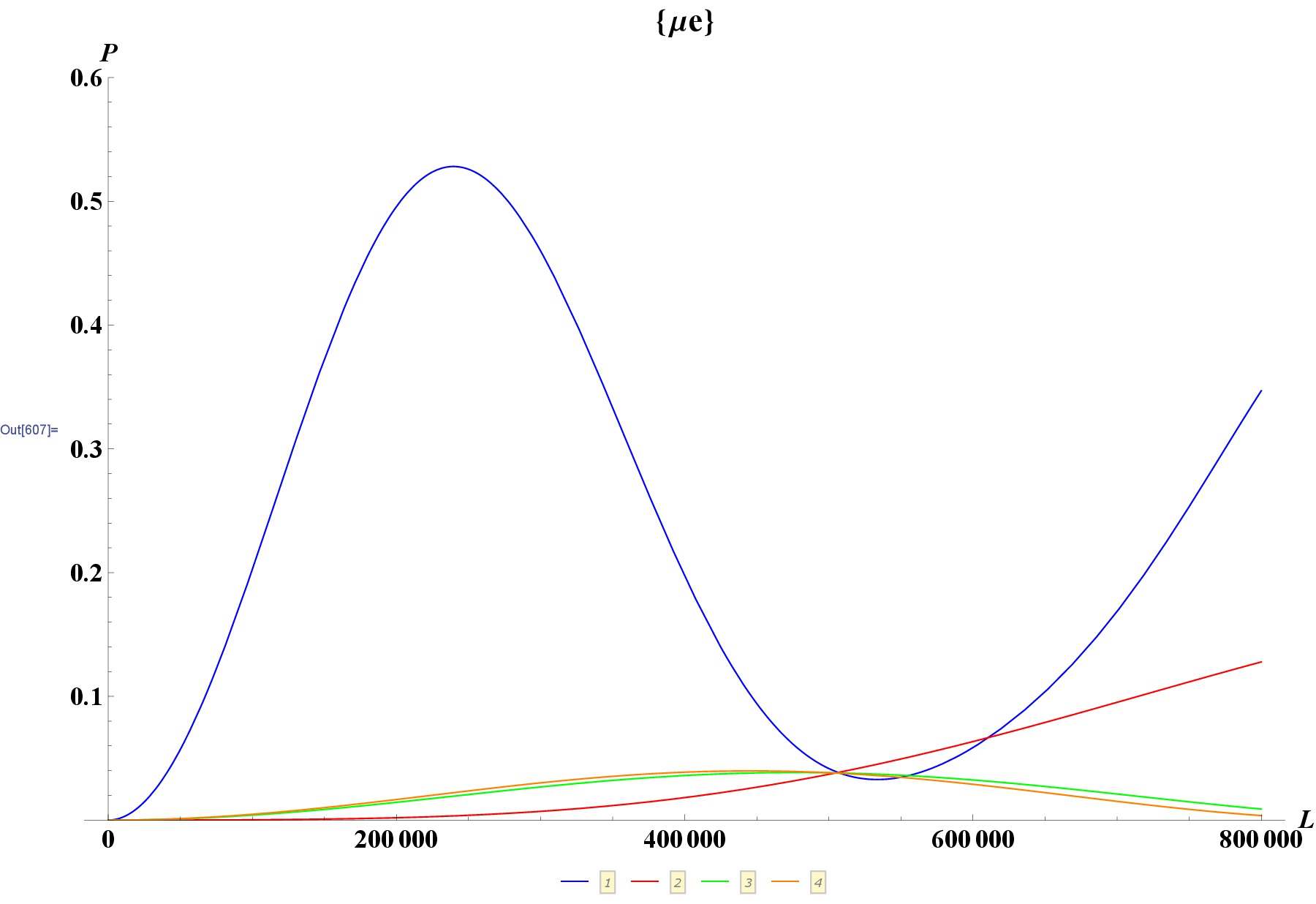}
\caption{Comparison of the $P_{\nu_{\mu},\nu_e}$ oscillation probability, as a function of the baseline L, for neutrino energy ${\rm E = 1 \, TeV}$, for
a ``standard theory", preserving Lorentz Invariance (orange curve) and for different versions of models including LIV, with parameters equal, respectively,
to $\delta f_{32} = \delta f_{21} = 4.5 \times 10^{-27}$ (blue), $\delta f_{32} = \delta f_{21} = 4.5 \times 10^{-28}$ (red) and $\delta f_{32} = \delta f_{21} = 4.5 \times 10^{-29}$ (green curve).}
\label{Pmue1tev3par}
\end{figure}

\begin{figure}[h]
\includegraphics[width=85mm]{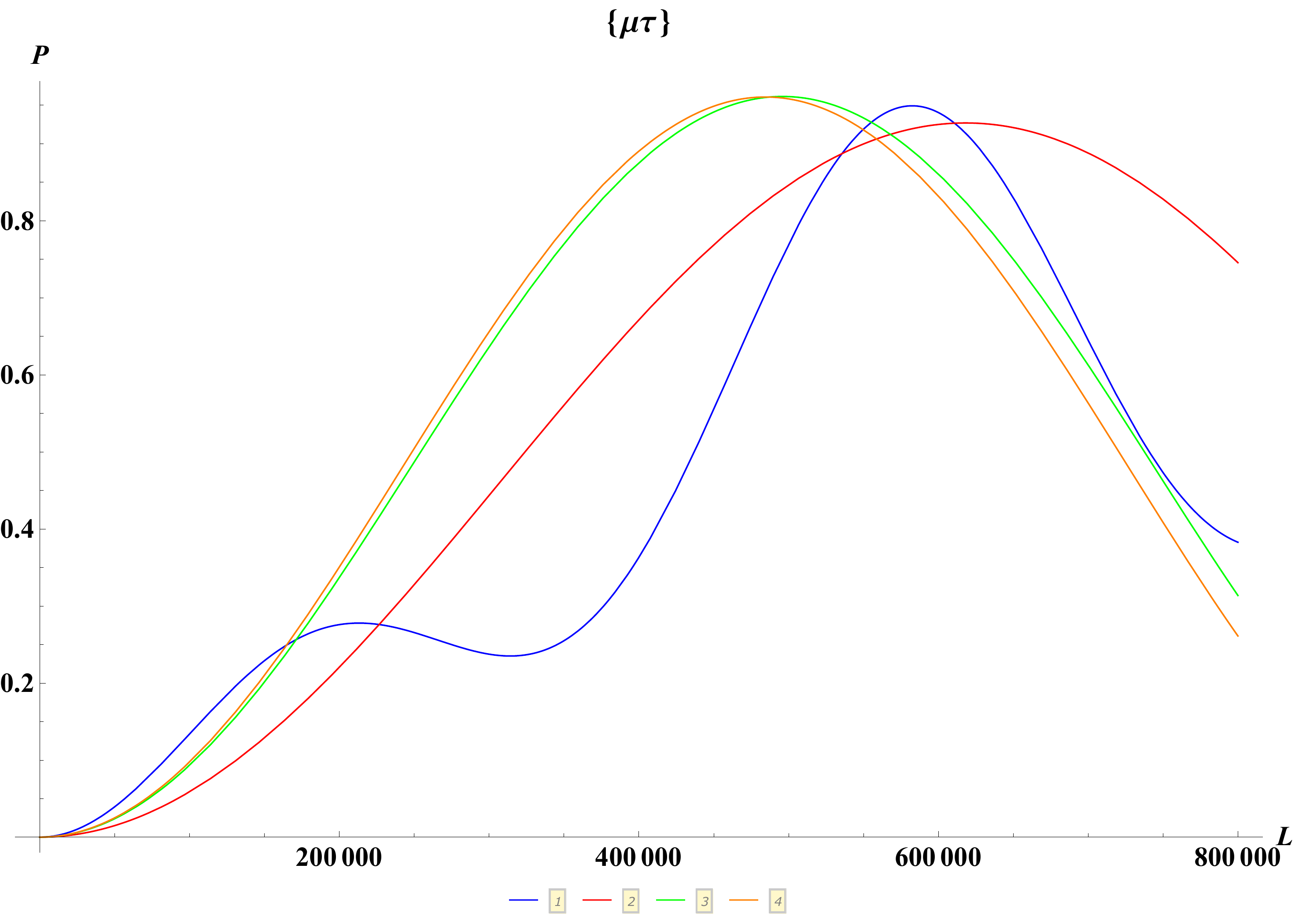}
\caption{Same analysis of fig.\ref{Pmue1tev3par}, but for the case of $P_{\nu_{\mu},\nu_{\tau}}$.}
\label{Pmutau1tev3par}
\end{figure}

\begin{figure}[h]
\includegraphics[width=90mm]{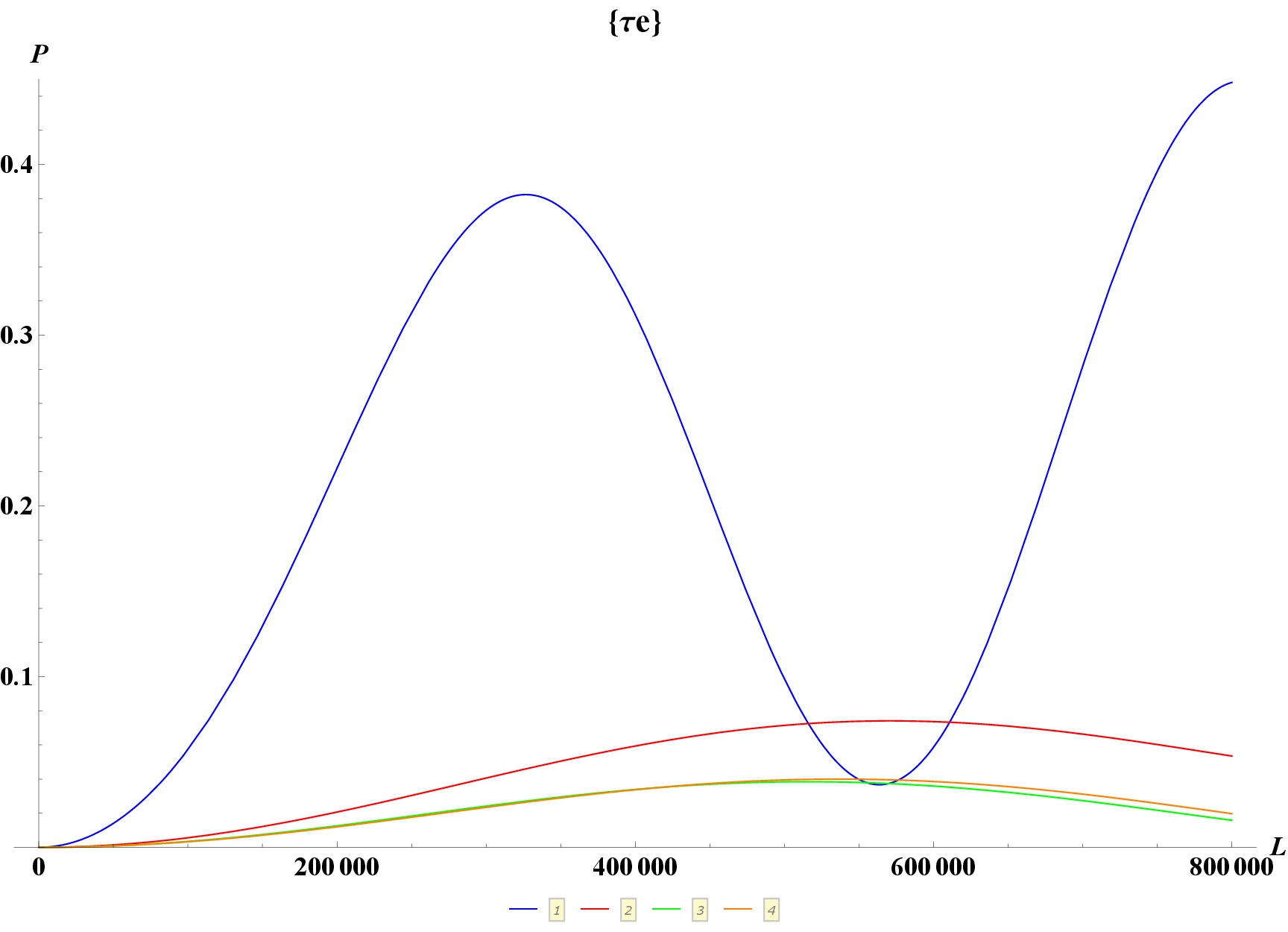}
\caption{Same analysis of fig.\ref{Pmue1tev3par}, but for $P_{\nu_{e},\nu_{\tau}}$}
\label{Petau1tev3par}
\end{figure}

\begin{figure}[h]
\includegraphics[width=90mm]{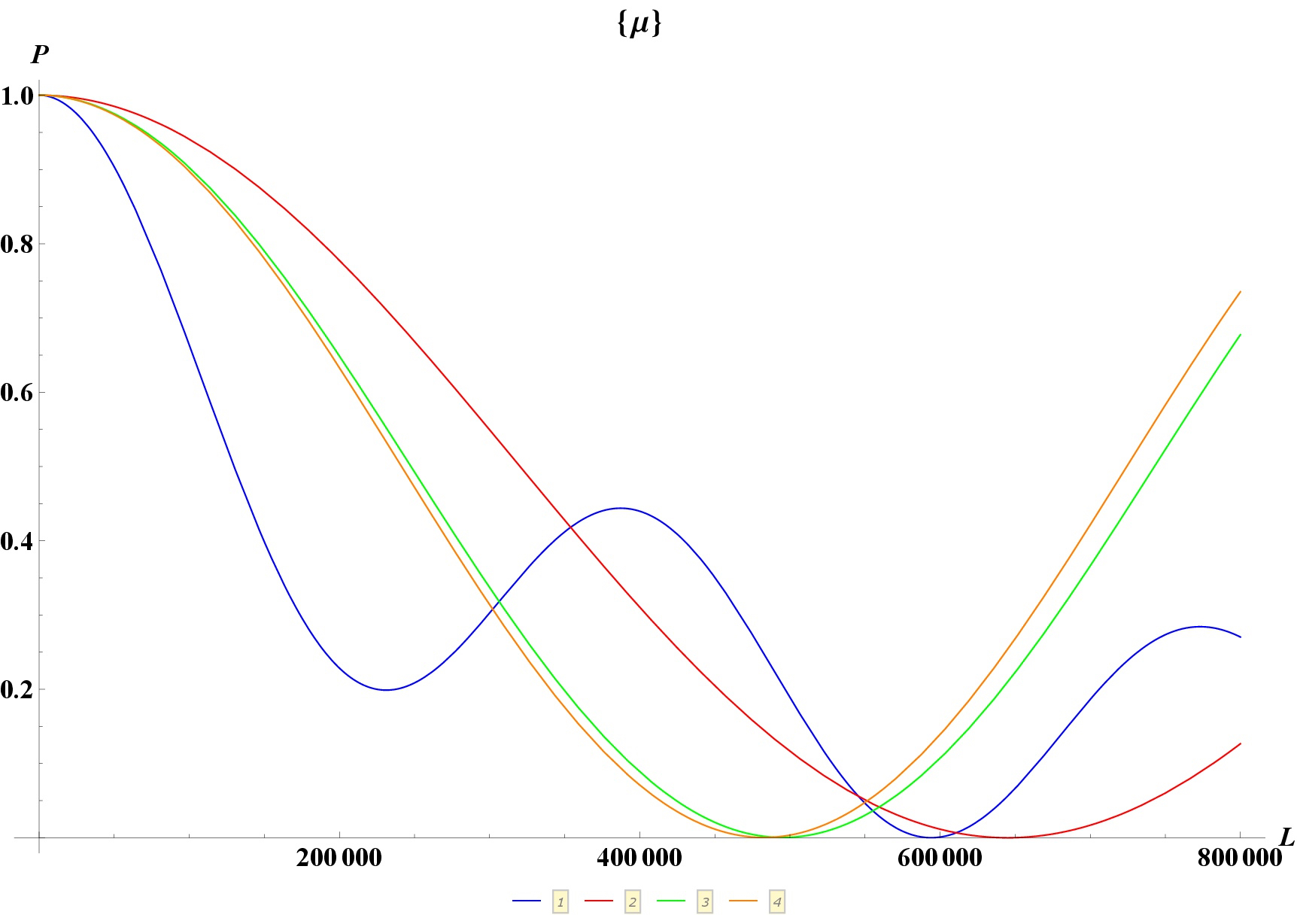}
\caption{Comparison of the results for the total muonic neutrino survival probability in a theory without LIV and in models with
LIV corrections, corresponding to three different values of the $\delta f_{kj}$ parameters, as illustrated in fig.\ref{Pmue1tev3par}.
Also the color code is the same adopted in figs.\ref{Pmue1tev3par}-\ref{Petau1tev3par}.}
\label{Pmumu1tev3par}
\end{figure}

The impact of the LIV corrections increases if one considers higher energy neutrinos. One can think, for instance, of neutrino energies in the region from TeV to PeV, interesting for present and future neutrino telescopes like  ANTARES~\cite{ANTARES}, KM3NET~\cite{KM3NET} and (for the higher energies mainly) IceCube~\cite{IceCube} (as analized  also in~\cite{Diaz:2013wia}). The analysis could be extended even to higher energies, as in the case of Ultra High Energy (above EeV) cosmic neutrinos, which are investigated, for instance, by Auger~\cite{Auger} and which will play a more and more relevant role in future in a multimessenger approach, further stimulated by the recent discovery of gravitational waves~\cite{Gravitational-waves}.
 Starting from the analysis of the ``lower" energies of this part of the spectrum, we analyzed the effect of Lorentz violation for a 1 TeV neutrino, considering 3 different sets of possible values for the $\delta f_{kj}$ parameters: in the first case we assumed $\delta f_{32} = \delta f_{21} = 4.5 \times 10^{-27}$ (corresponding to the present limit derived by SuperKamiokande), while in the other 2 cases we explored values of the $\delta f_{kj}$ lower, respectively, of one and two orders of magnitude.
The results, reported in the series of graphs of figs.\ref{Pmue1tev3par}-\ref{Petau1tev3par} for the 3 oscillation probabilities
($P_{\nu_{\mu},\nu_{e}}$, $P_{\nu_{\mu},\nu_{\tau}}$, $P_{\nu_{e},\nu_{\tau}}$) and in fig.\ref{Pmumu1tev3par} for the total survival probability of muonic neutrino, are promising. It is evident that the curves corresponding to the LIV expressions obtained for $\delta f_{32} = \delta f_{21} = 4.5 \times 10^{-27}$ (blue lines) are signiflcantly different from the ones obtained in absence of LIV violations (orange curves). Moreover, the corrections due to LIV remain significant also for $\delta f_{kj}$ parameters one order of magnitude lower (red) and they are in any case apprecciable even for $\delta f_{32} = \delta f_{21} = 4.5 \times 10^{-29}$ (green curve), at least for values of the baseline sufficiently high (L above 400000 km in the first oscillation cycle).
Hence, there is the hope that, by selecting the appropriate experimental context, in future one could use the detailed study of high energy neutrinos to further constraint the values of the coefficients controlling the possible sources of Lorentz Invariance Violation.

In a full phenomenological analysis of any realistic experimental situation the information about the oscillation probability
is central, but, obviously, it  must be complemented by an accurate knowledge of the expected fluxes in absence of oscillation for every flavor neutrinos and of the different interaction cross sections.
The number $N_{\alpha,\beta}$ of detected transition events due to the $\nu_{\alpha} \to \nu_{\beta}$ flavor oscillation will be given, as a function of the energy E of the neutrinos that underwent the oscillation and of the distance L they travelled from the production to the detection point, by:\\
\begin{equation}
{\rm N_{\alpha,\beta} \propto \Phi_{\alpha} (L, E)  \, P_{\nu_{\alpha},\nu_{\beta}} (L, E) \, \sigma_{\beta} (E)} \, ,
\nonumber
\end{equation}
where $\Phi_{\alpha}$ and $\sigma_{\beta}$ represent, respectively, the predicted flux of neutrinos of flavor $\alpha$ in absence of oscillation and the cross section of interaction for the $\nu_{\beta}$ neutrinos with the detector, which depends upon the specific
experiment studied. In most cases this information must be integrated over the neutrino energies (and eventually also over the distances L, that in many experiments are translated into angular bins) and the integrals has to be convoluted with functions describing the detector resolution and efficiencies.
From the comparison, by means of statistical methods, between the experimental results and the theoretical predictions one can extract the information about the impact of eventual LIV violations present in the model or put constraints on the order of magnitude of the coefficients ruling these effects.
We are performing such an analysis for different experimental situation of particular interest and the work describing the results of
our study is in progress\cite{noifuturo}.

\section{Conclusions}

Lorentz covariance is one of the fundamental properties of space time in the standard version of relativity.
Nevertheless, the possibility of small violations of this fundamental invariance has been explored in different extensions of the Standard Model and more
generally in many exotic theories and a variety of possibile experiments searching for signals of LIV (Lorentz invariance violations) have been proposed
over the years. A significant numbers of these tests has to do with the study of neutrino properties, also because neutrino phenomenology is extremely
reach and spans over a very wide range of energies.

In this paper we consider a class of models, more widely discussed in~\cite{Torri}, in which the possibility of LIV is introduced starting from a modified version
of dispersion relations and is founded on a more general geometrical description, making use of Finsler geometry. The choice of the particular form of the terms
violating Lorentz Invariance, represented as an homogeneous function of $\frac{|\overrightarrow{p}|}{E}$, guarantees the possibility of  preserving a metric structure and, moreover, the LIV corrections are chosen in such a way to respect the isotropy of space time and the CPT invariance. The effect of the perturbative LIV corrections we introduce in our model is that of modifying the kinematics, without changing the degrees of freedom of the theory and the interactions and preserving the internal $SU(3)\times SU(2)\times U(1)$ symmetry. The kind of model we obtain can be considered an extended version of the Standard Model, equivalent to models studied for instance in~\cite{Koste2}, in the case in which one restricts the LIV to CPT even terms.

In this paper we analyze the impact of LIV perturbative corrections present in our model on neutrino phenomenology, both in an hamiltonian approach
and by means of a detailed study of the oscillation probabilities, that rule the different possible flavor transitions. The modification of the dispersion relations
with the introduction of Lorentz invariance violating terms (that can be treated with a sort of perturbative approach) imply a change in the form of the
``phase differences'' $\Delta \phi_{ij}$, which enter  as argument of $sin^2 (\Delta \phi_{ij})$, representing the contribution of the $i, j$ mass eigenstates
to the oscillation probability functions. As shown in eq.(~\ref{diffphase}), in addition to the usual term $\frac{\Delta m_{ij}^2 L}{E}$, another contribution appears
in the expression for $\Delta \phi_{ij}$, dependent upon the differences between the LIV coefficients for the different mass generations ($\delta f_{ij}$) and
proportional to $L \times E$. This means that in our model the presence of LIV has an impact on the neutrino oscillation only if these terms are not identical
for all the mass generations. Besides, the fact that the LIV corrections are proportional to E, instead of $\frac{1}{E}$, implies that, in order to be consistent
with the data from the different oscillation experiments, these corrections must represent small perturbations which do not change the general
``pattern" of neutrino oscillation. Nevertheless, these corrections could be significant in particular experimental situations and with an appropriate choice
of the experimental tests it could be possible to further constrain the possible values of the LIV coefficients.
It is important to underline that this kind of oscillation corrections are foreseen even by other approaches, based on EFT, such as in~\cite{Liberati2}. Our model faculty of reproducing these predictions constitutes a test of validity of our geometrical approach, which gives a theoretical background to the introduction of MDRs, in a context that can be considered a different ``philosophical approach''.

We deeply investigated the impact of these LIV corrections, comparing all the different oscillation probabilities evaluated in our model in presence of LIV with the analogous expressions in absence of LIV, for different fixed values of neutrino energy (selected in such a way to cover different energy regions of phenomenological interest) and spanning over a wide range of values for the baseline between the neutrino production and detection points, using the complete oscillation theory with all the mass eigenstates.

We showed that significant deviations from the ``standard'' values of oscillation probabilities could be present already for energies around 1 GeV
if one assumes values for the LIV coefficients of the same order of magnitude usually considered in literature~\cite{altri-fenomenologici} (around $10^{-23}$).
On the other hand, if one limits significantly the magnitude of LIV corrections, considering the values recovered in a recent analysis~\cite{SK-test-LIV} by
SuperKamiokande collaboration for the CPT even LIV coefficients, the effect of LIV on the oscillation probabilities starts to become evident for higher neutrino energies (around 100 GeV).

We studied in detail the situation for 1 TeV neutrinos, analyzing the improvement that, in this case, should
be possibile to obtain on the limits for the LIV coefficients, and we also discussed the scenarios that could be even more promising of the
future studies of ultra high energy neutrinos (like the cosmic ones).
A series of real possible experimental situations, corresponding to various neutrino sources of different energies for present and future experiments, are presently
under investigation and will be discussed in a separate work~\cite{noifuturo}.\\

\end{document}